\documentclass[iop,numberedappendix]{emulateapj}
\slugcomment{Accepted by \apj} 
\usepackage{natbib}
\usepackage{amsmath}
\usepackage{booktabs}
\usepackage{longtable}
\usepackage{color}
\usepackage{graphicx, subfigure}
\usepackage{hyperref}
\usepackage{enumitem}%remove indent from enumerate
\newcommand{\mr}{\mathrm} 
\newcommand{\di}{\mathrm{d}} 
 
\newcommand{\St}{\mathrm{St}} 
\newcommand{\Rey}{\mathrm{Re}} 

\newcommand{\rom}[1]{%
  \textup{\uppercase\expandafter{\romannumeral#1}}%
}

\defcitealias{OC2007}{OC2007}
\defcitealias{dust1}{paperI}

\begin{document}
\title{Impact of magneto-rotational instability on grain growth 
in protoplanetary disks: 
II. Increased grain collisional velocities}
\author{Munan Gong\altaffilmark{1}, Alexei V. Ivlev\altaffilmark{1},
Vitaly Akimkin\altaffilmark{2}, and Paola Caselli\altaffilmark{1}}
\altaffiltext{1}{Max-Planck Institute for Extraterrestrial Physics,
Garching by Munich, 85748, Germany; munan@mpe.mpg.de}
\altaffiltext{2}{Institute of Astronomy, Russian Academy of Sciences,
Pyatnitskaya str. 48, Moscow, 119017, Russia}

\begin{abstract}
Turbulence is the dominant source of collisional velocities for grains with a wide range of sizes in protoplanetary disks. So far, only Kolmogorov turbulence has been considered for calculating grain collisional velocities, despite the evidence that turbulence in protoplanetary disks may be non-Kolmogorov. In this work, we present calculations of grain collisional velocities for arbitrary turbulence models characterized by power-law spectra and determined by three dimensionless parameters: the slope of the kinetic energy spectrum, the slope of the auto-correlation time, and the Reynolds number. 
%----This can be deleted if word limit is a concern
%We calculate the grain collisional velocities using numerical integration, and derive simple and accurate analytic approximations. We provide physical understanding of the collisional velocities, as well as scaling relations with the Stokes numbers and turbulence properties. 
%Our Python scripts for calculating grain collisional velocities were made publicly available.
%---------
The implications of our results are illustrated by numerical simulations of the grain size evolution for different turbulence models.
We find that for the modeled cases of the Iroshnikov-Kraichnan turbulence and the turbulence induced by the magneto-rotational instabilities, collisional velocities of small grains are much larger than those for the standard Kolmogorov turbulence. This leads to faster grain coagulation in the outer regions of protoplanetary disks, resulting in rapid increase of dust opacity in mm-wavelength and possibly promoting planet formation in very young disks.
\end{abstract}

\section{Introduction}
Interstellar dust grains plays an important role in many aspects of
astrophysics: they are building blocks of planets, commonly used gas tracer,
and catalyst of molecular chemistry. 
All these processes depend on the
size distribution of dust grains, and great efforts have been made to model the
process of grain growth, especially in protoplanetary disks 
\citep[see reviews by][]{BM2008, Testi2014, Birnstiel2016}.
In various astrophysical environments, turbulence is the major driving force for grain
growth. Turbulent motions stir up the grains, leading to their mutual collisions.
In protoplanetary disks, for example, turbulence is among the
dominant sources for collisional velocities between grains in the size
range of microns to meters \citep{Birnstiel2011}.

The calculation of grain collisional velocity induced by turbulence generally
relies on one critical assumption: the turbulence is Kolmogorov with a kinetic
energy spectrum of $E(k)\propto k^{-5/3}$.
\citet{Voelk1980} and \citet{Markiewicz1991} made the ground-laying work of
calculating grain collisional velocities in Kolmogorov turbulence. Later 
\citet[][hereafter \citetalias{OC2007}]{OC2007} derived the analytic
expressions for grain collisional velocities in Kolmogorov turbulence, which
were soon adopted in many grain coagulation codes
\citep[e.g.][]{DH2008, Okuzumi2012, Akimkin2020}. 
The grain collisional velocities derived from these analytic 
V\"{o}lk-type models were tested by direct numerical simulations in \citet{PP2015, Ishihara2018} and \citet{Sakurai2021}. 
The simulations showed that while V\"olk-type models suffer from several drawbacks, such as the neglect of turbulent clustering of same-size grains (which enhances their collisional rates) and a reduction of the rms collisional velocity (which reduce the collisional rates), the overall grain collisional velocities derived from V\"olk-type models are still accurate within a factor of $\sim 2$.
With its relative accuracy and simplicity, the formulae in \citetalias{OC2007} remain the standard adopted in the current literature.

The astrophysical turbulence, however, does not necessarily have the Kolmogorov spectrum.
The presence of magnetic fields is expected to change the turbulence cascade, and
many alternative theories have been proposed to describe the magneto-hydrodynamic
(MHD) turbulence. The Iroshnikov-Kraichnan (IK) theory, for example, predicts
$E(k) \propto k^{-3/2}$ \citep{Iroshnikov1964, Kraichnan1965}. Alternatively, the 
Goldreich-Sridhar theory predicts $E(k_\parallel) \propto
k_\parallel^{-2}$ parallel to the mean magnetic field, and 
$E(k_\perp) \propto k_\perp^{-5/3}$ perpendicular to the mean magnetic field
\citep{GS1995}. These theories, however, assume that there is a dominant mean
magnetic field. In many astrophysical environments such as the protoplanetary
disks, the magnetic field is weak and the mean field varies on spatial and
temporal scales of the turbulent cascade. It is unclear whether the 
theoretical predictions by the IK or Goldreich-Sridhar turbulence models 
still hold in these environments.

In our previous paper \citet[][hereafter \citetalias{dust1}]{dust1}, we
preformed numerical simulations of the MHD turbulence in protoplanetary
disks generated by the magneto-rotational instabilities (MRI). 
We observed a persistent kinetic energy spectrum of $k^{-4/3}$, 
which appears to be converged 
in terms of numerical resolution. 
This $k^{-4/3}$ energy spectrum has also been observed in  many other MHD
turbulence simulations in the literature (see Table 6 in \citetalias{dust1}).
To further investigate this phenomenon, we also performed driven turbulence
simulations with and without the magnetic field and
obtained the same energy spectrum. We concluded that the $k^{-4/3}$ power-law slope
is likely due to the bottleneck effect near the dissipation scale of the turbulence \citep{Ishihara2016}.
Due to the limited numerical resolution, we were not able to constrain whether the
$k^{-4/3}$ energy spectrum extends to a larger dynamical range. In addition, we
found the turbulence auto-correlation time to vary close to $\propto k^{-1}$,
which is steeper than that for the Kolmogorov turbulence. 
Moreover, the injection scale of the MRI
turbulence is determined by the fastest growing mode of the MRI -- not by the
scale-height of the disk assumed as the injection scale in \citetalias{OC2007}. 
All these factors -- 
the energy spectrum, the auto-correlation time and the injection scale of the
turbulence -- can have a big impact on the grain collisional velocities.

Recently, \citet{Grete2020} performed numerical simulations of weakly
magnetized MHD turbulence, and found the same kinetic energy spectrum slope of
$k^{-4/3}$. They analysed the energy transfer mechanisms in their simulations, and
argued that magnetic tension must be the dominant force for energy transfer across
scales.  The energy transfer mechanism in Kolmogorov turbulence, the
kinetic energy cascade, is suppressed in this case by the magnetic tension, to which they
attributed the cause of the shallower $k^{-4/3}$ energy spectrum. Although they
believe that the bottle-neck effect is not the cause of the $k^{-4/3}$ slope, their
numerical resolution is still limited at $(2048)^3$. The power-law slope
measured in \citet{Grete2020} spans within $\sim 2$ dex of the dissipation scale, where the
bottle-neck effect is known to affect the energy spectrum \citep{Ishihara2016}. 
Moreover, there is
no theoretical understanding so far about why the energy transfer by magnetic
tension force may lead to the $k^{-4/3}$ energy spectrum. However, if the energy
spectrum is indeed determined by the magnetic tension, the $k^{-4/3}$ slope can
represent the inertial range of the turbulence cascade and extend to a much
wider dynamic range far from the dissipation scale. 

In addition, pure hydrodynamic instabilities can also generate turbulence in 
the protoplanetary disks in the absence of magnetic fields. For example, the 
subcritical baroclinic instability (SBI) driven by the radial entropy gradient 
\citep{KB2003, Klahr2004, Peterson2007, LP2010},  
and the vertical shear instability (VSI) driven by the strong vertical shear 
\citep{Nelson2013, SK2016} can both generate long-lived turbulence in disks. 
Numerical simulations have found that the turbulence 
induced by the SBI or VSI can have much 
steeper kinetic energy spectra than the Kolmogorov turbulence across certain scales 
\citep{KB2003, Manger2020}.

Given the uncertainty in the turbulence properties, this paper aims to provide
insights into how non-Kolmogorov turbulence may affect grain collisional
velocities and grain growth. In Section
\ref{section:method}, we describe the turbulence models and the procedure
for calculating the grain collisional velocities. We focus on three examples: the
Kolmogorov turbulence, the IK turbulence, and the MRI turbulence described in
\citetalias{dust1}. The results are shown in Section \ref{section:results}.
Section \ref{section:analytic_approximation} derives the analytic approximation
for the collisional velocities assuming a general case of power-law turbulence spectrum,
and Section \ref{section:comparision_numerical} 
compares the analytic approximation with accurate numerical
integration. We also supply publicly available Python scripts that implemented
our formulae for calculating the collisional velocities.
Section \ref{section:limiting_behaviors} shows the dependence of grain collisional velocity on turbulence
parameters, by describing its behavior in
different limiting regimes. Section \ref{section:simulation} presents an
application of our work: we calculate grain growth in protoplanetary disks with
different turbulence models, and estimate the fragmentation and drift barrier for grain growth due to non-Kolmogorov turbulence. Finally, Section \ref{section:summary} gives a summary of this work.

\section{Method}\label{section:method}
We follow the method in \citet{Voelk1980}, \citet{Markiewicz1991} 
and \citetalias{OC2007} to calculate grain collisional velocities.
These previous works considered only Kolmogorov turbulence. Here we 
generalize to a generic turbulence model with arbitrary power-law 
slopes of energy spectrum and auto-correlation time. We first describe the turbulence
model, and then the steps to calculate the turbulence-induced grain collisional
velocities. For the convenience of the reader, we summarize the important
notations used in this paper in Table \ref{table:notations}.
\begin{table}[htbp]
    \centering
    \caption{Summary of notations for the key physical variables}
    \label{table:notations}
    \begin{tabular}{cc}
        \tableline
        \tableline
        Symbol &Meaning\\
        \tableline
        %$\mathbf{v}_K$  &Keplerian velocity\\
        $\mathbf{v}$    &gas velocity\\
        $\delta \mathbf{v}$ & $\mathbf{v} - \mathbf{v}_K$, turbulent gas velocity\\
        $v_\mathrm{tot}$ &large-scale turbulent velocity (Eq. (\ref{eq:vt}))\\
        $v_\mathrm{rel}(k)$ &relative velocity between grain and eddy
                         (Eq. (\ref{eq:vrel}))\\
        $v_p\equiv \delta v_p$ &turbulence induced grain velocity (Eq. (\ref{eq:tauf}))\\
        $\Delta v_{12}$ &collisional velocity 
              between grain 1 and 2 (Eq. (\ref{eq:v12}))\\
        $\St$ &Stokes number (Eq. (\ref{eq:St}))\\
        $\Rey$ &Reynolds number (Eq. (\ref{eq:re}))\\
        $\tau_f$ &grain friction/stopping time (Eq. (\ref{eq:tauf}))\\
        $\tau_\mathrm{cross}(k)$ &eddy crossing time (Eq. (\ref{eq:tau_cross}))\\
        $\tau(k)$ &eddy auto-correlation time (Eq. (\ref{eq:tauk})) \\
        $E(k)$& kinetic energy spectrum (Eq. (\ref{eq:Ek}))\\
        $k_L$ &injection scale (Fig. \ref{fig:Ek_tauk})\\
        $k_\eta$ &dissipation scale (Fig. \ref{fig:Ek_tauk})\\
        $p$ &power-law slope of $E(k)$\\
        $m$ &power-law slope of $\tau(k)$\\
        \tableline
        \tableline
    \end{tabular}
\end{table}

\subsection{Turbulence Model}
There are two important properties of turbulence that determine the grain
collisional velocities, the kinetic energy spectrum $E(k)$ and the eddy
auto-correlation time $\tau(k)$. 
We assume the turbulence has a kinetic energy spectrum 
\begin{equation}\label{eq:Ek}
    E(k) = E_L \left( \frac{k}{k_L} \right)^{-p},\quad k_L \leq k \leq k_\eta,
\end{equation}
where $k_L$ and $k_\eta$ are the injection scale and dissipation scale of
the turbulence. Outside of the range $k_L \leq k \leq k_\eta$,
we simply assume $E(k)=0$.\footnote{In \citetalias{dust1}, we found that the injection
scale $k_L$ is similar to the fastest growing mode of the MRI in the disk,
$k_\mr{MRI}$. At $k < k_\mr{MRI}$, there is still a region
with $E(k) > 0$. However, because the slope of $E(k)$ in this region is much
shallower than at $k > k_\mr{MRI}$, the kinetic energy is dominated by $k\approx k_L$.
Therefore, by using the simple assumption of $E(k)=0$ for $k<k_L$, the
dust collisional velocities are not affected significantly.}

The dissipation scale is determined by the Reynolds number, 
\begin{equation}\label{eq:re}
    \Rey = \frac{vL}{\nu},
\end{equation}
where $\nu$ is the viscosity, $v$ is the velocity, and $L$ is the length scale. Usually, the Reynolds number is defined for the
largest turbulence eddy, $\Rey = \Rey(k_L) = v(k_L)L(k_L)/\nu$.
For turbulence eddy $k$, $v(k) = \sqrt{k E(k)}
\propto k^{(1-p)/2}$, and $L(k)\propto 1/k$. At the dissipation scale,
$\Rey(k_\eta)=v(k_\eta)L(k_\eta)/\nu=1$. This gives,
\begin{equation}
    \Rey = \frac{v(k_L)L(k_L)}{v(k_\eta)L(k_\eta)} 
    = \left( \frac{k_\eta}{k_L} \right)^{(1+p)/2},
\end{equation}
or equivalently,
\begin{equation}
    \frac{k_\eta}{k_L} = \Rey^{2/(1+p)}.
\end{equation}

The large-scale (total) turbulent velocity $v_\mathrm{tot}$ is defined as
\begin{equation}\label{eq:vt}
    v_\mr{tot}^2 = \int \di k E(k).
\end{equation}
From the Plancherel theorem, $v_\mathrm{tot}^2 = \langle v^2 \rangle$,
where $v$ is the magnitude of turbulent gas velocity and ``$\langle ~
\rangle$'' denotes the spacial average.
Integrating Equation (\ref{eq:vt}), we have
\begin{equation}\label{eq:vt_int}
    v_\mr{tot}^2 = \frac{E_L k_L}{p-1}\left[ 
    1 - \left( \frac{k_\eta}{k_L} \right)^{-(p-1)} \right]
    \approx \frac{E_L k_L}{p-1}.
\end{equation}
For the integration to converge, it requires $p>1$.

\begin{figure*}[htbp]
\centering
\includegraphics[width=\linewidth]{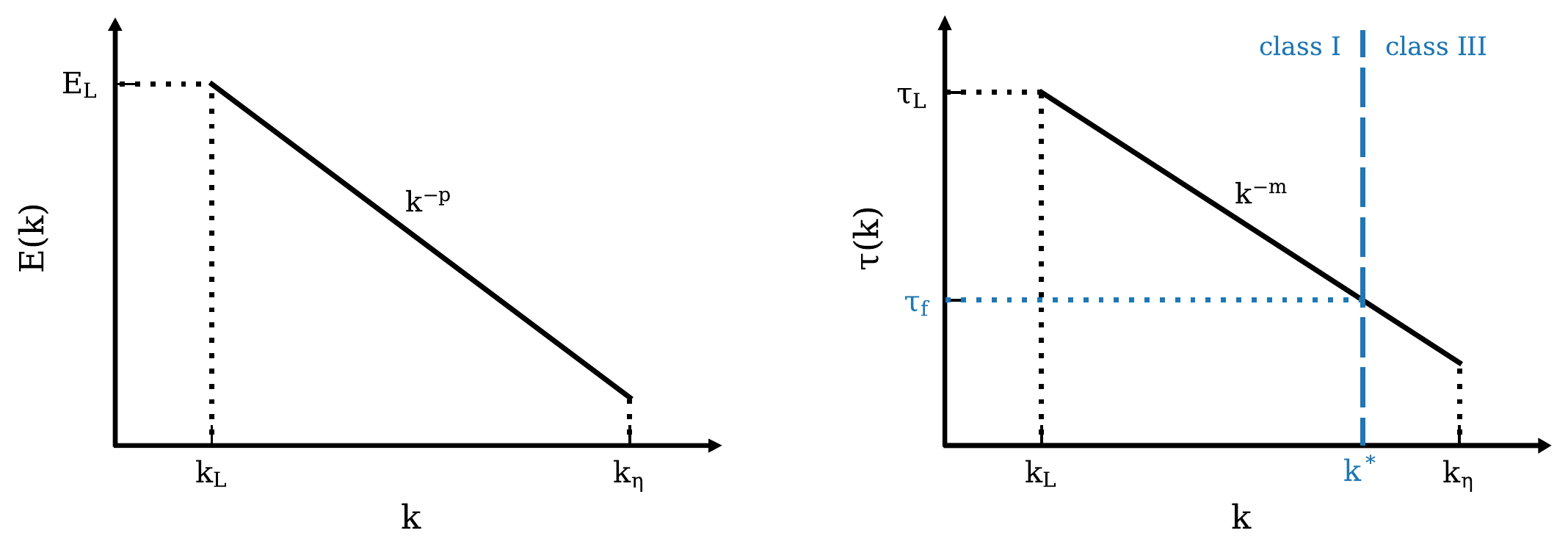}
    \caption{Schematic diagrams of the kinetic energy spectrum $E(k)$ (left
    panel) and the eddy auto-correlation time $\tau(k)$ (right panel).
    $k_L$ and $k_\eta$ denote the injection and dissipation scales respectively.
    The blue annotations mark the transition between class \rom{1} and class \rom{3} eddies (see Equations (\ref{eq:eddy_class}) - (\ref{eq:kstar_st}))
    \label{fig:Ek_tauk}}
\end{figure*}

The corresponding turbulent auto-correlation time in
the inertial range is 
\begin{equation}\label{eq:tauk}
    \tau(k) = \tau_L \left( \frac{k}{k_L} \right)^{-m}.
\end{equation}
Figure \ref{fig:Ek_tauk} illustrates the models for $E(k)$ and $\tau(k)$. 
For the detailed definitions of $E(k)$ and $\tau(k)$ see \citetalias{dust1}.

\citet{OC2007} assumed $\tau(k) = 1/(k\sqrt{2kE(k)})$ from the kinetic cascade,
which gives $\tau_L=1/(k_L\sqrt{E_L k_L})$ and $m=(3-p)/2$. In \citetalias{dust1}, we did
observe that $\tau(k) \approx 1/(k\sqrt{2kE(k)})$ for the MRI as well as driven
turbulence.\footnote{In \citetalias{dust1}, we found that $\tau(k) \approx \min
\{ 1/\Omega, 1/(k\sqrt{2kE(k)})\}$ for the MRI turbulence,
where $\Omega$ is the local orbital frequency. With $k_\mr{MRI}\approx \Omega /
\langle v_{A,z} \rangle$, where $\langle v_{A,z} \rangle$ is the average Alfven
speed in the vertical direction, and $v_\mr{tot}\approx \sqrt{E_L k_\mr{MRI}}
\approx \langle v_{A,z} \rangle$ (turbulent velocity comparable to the Alfven
speed), we have $(k_\mr{MRI}\sqrt{2k_\mr{MRI}E(k_\mr{MRI})} \approx
\Omega$. This means $\tau(k) \approx 1/(k\sqrt{2kE(k)})$ at $k>k_\mr{MRI}$. } 
%\munan{To Alexei: with the argument of interactive Alfven wave packets, IK turbulence
%would have m=1/2. Any experiment/simulation confirming any of this?}
In principle, $\tau(k)$ can be influenced also by other physical processes
such as the interaction between the gas and the magnetic field. Without
losing generality, we keep $p$ and $m$ as separate parameters. We focus on three
turbulence models shown in Table \ref{table:turb_model}, the Kolmogorov
turbulence, the IK turbulence and the MRI turbulence in \citetalias{dust1}.
We note that our method can also be applied to other turbulence models with 
arbitrary values of $p$ and $m$.

\begin{table}[htbp]
    \centering
    \caption{Parameters for turbulence models considered in this work. }
    \label{table:turb_model}
    \begin{tabular}{l cccc}
        \tableline
        \tableline
        model\tablenotemark{a} &$p$  &$m$ &$(p-1)/m$\\
        \tableline
        Kolmogorov         &5/3   &2/3   &1    \\
        IK                 &3/2   &3/4   &2/3 \\
        \citetalias{dust1} &4/3   &5/6   &2/5 \\
        \tableline
        \tableline
    \end{tabular}
    \tablenotetext{1}{The parameter $(p-1)/m$ determines the scaling between the collisional velocity and the Stokes number for small grains (Equation (\ref{eq:v12_St})). }
\end{table}

\subsection{Turbulence-induced Collisional Velocities\label{section:method:vcoll}}
The dynamical property of a dust grain is characterized by 
its friction time (also often called the
stopping time) $\tau_f$. The randomly fluctuating component of grain 
velocity $\delta \mathbf{v}_p$ follows \citep[Equation (5) in][]{Voelk1980}: 
\begin{equation}\label{eq:tauf}
    \frac{\di \delta \mathbf{v}_p}{\di t} 
    = \frac{\delta \mathbf{v} - \delta \mathbf{v}_p}{\tau_f},
\end{equation}
where $\delta \mathbf{v}$ is the random component of the gas velocity, such as
the turbulent velocity in the protoplanetary disk.
For the simplicity of notations, we hereafter drop the $\delta$ 
in Equation (\ref{eq:tauf}), and use $\mathbf{v}_p$ to denote the randomly
fluctuating component of grain velocity induced by turbulence. 

It is convenient to define the dimensionless Stokes number
\begin{equation}\label{eq:St}
\St = \tau_f / \tau_L.
\end{equation}
Physically, $\tau_f$ or $\mathrm{St}$ is determined by the properties of both the
grain and gas, and the relative velocity between them \citep{Youdin2010}. For
spherical grain in the Epstein drag regime \citep{Epstein1924}, 
the friction time $\tau_f = \rho_s a/(\rho_g v_\mr{th})$, where $\rho_s$ and $a$ are
the material density and radius of the grain, $\rho_g$ the gas density, and $v_\mr{th}$ the mean thermal velocity of the gas. In a protoplanetary disk, the Stokes number can be written as \citep{Birnstiel2016},
\begin{equation}
    \St = \frac{\pi}{2} \frac{\rho_s a}{\Sigma_g},
\end{equation}
where $\Sigma_g$ is the gas surface density. 
Using the minimum mass solar
nebular model (MMSN) in \citet{Hayashi1981}, $\St\approx 10^{-8} - 1$ for
typical grain sizes from $0.1~\mathrm{\mu m}$ to $1~\mathrm{cm}$ at
$1-100~\mathrm{AU}$.

In this work, only perturbations from the gas
motions on dust gains is considered, and the back reaction of dust
grains onto the gas is ignored.

For two dust grains with friction times $\tau_{f1}$ and $\tau_{f2}$, their
collisional velocity $\Delta v_{12}$ can be obtained by following the steps below:
\begin{enumerate}[leftmargin=*]
\item Calculate the relative velocity $v_\mathrm{rel}(k)$ between 
    the eddy $k$ and each dust grain 
     \citepalias[see Equation (19) in][]{OC2007}:
    \begin{equation}\label{eq:vrel}
        v_\mathrm{rel}^2(k) = v_\mathrm{sys}^2 + \int_{k_L}^{k} \di k' E(k')\left(
        \frac{\tau_f}{\tau_f + \tau(k')} \right)^2,
    \end{equation}
    where $v_\mathrm{sys}$ is the systematic velocity of the dust
    not driven by turbulence, such as the radial drift by pressure-gradient
    driven headwind or vertical settling due to the stellar gravity. Through out this work, we
    assume that the turbulent motions dominate and set $v_\mathrm{sys}=0$.
    %TODO: do we have to consider non-zero v_sys?
    
\item Determine the classes of eddies for each dust grain. The concept of
    ``eddy classes'' is first introduced by \citet{Voelk1980}. For a given dust
    grain with the friction time $\tau_f$ and a given eddy $k$, the eddy class
    is determined by 
    \begin{equation}\label{eq:eddy_class}
        \begin{cases}
            \tau_f < \mathrm{min}\{\tau(k), \tau_\mathrm{cross}(k)\},
                    & \text{class \rom{1} eddy}\\
            \tau_f \geq \mathrm{min}\{\tau(k), \tau_\mathrm{cross}(k)\}, 
                    & \text{class \rom{3} eddy},
        \end{cases}
    \end{equation}
    where
    \begin{equation}\label{eq:tau_cross}
        \tau_\mathrm{cross}(k)=\frac{1}{k v_\mathrm{rel}(k)}
    \end{equation}
    is the timescale
    on which the grain moves across the eddy. Grains are well-coupled with the class \rom{1} eddies. This corresponds to small grains, for which $\tau_f$ is
    short enough that the grain ``forgets'' its initial motion and moves with the
    gas before it leaves the eddy or the eddy decays. On the contrary,
    grains are only weakly coupled with the class \rom{3} eddies. Such grains are large
    enough that $\tau_f$ is long, and the eddy only exerts small
    perturbations to their motions. Because both $\tau(k)$ and 
    $\tau_\mathrm{cross}(k)$ increase with $k$, a grain is better coupled
    with the large eddies than the small eddies. The transition scale
    between class \rom{1} and class \rom{3} eddies is defined as $k^*$: the
    eddies are class \rom{1} for $k < k^*$ and class \rom{3} 
    for $k\geq k^*$. $k^*$ is a function of $\tau_f$ and can be solved by 
    \begin{equation}\label{eq:kstar}
        \mathrm{min}\{\tau(k^*), \tau_\mathrm{cross}(k^*)\} = \tau_f.
    \end{equation}
    Appendix \ref{appendix:class} shows that $k^*$ 
    can be approximated using $\tau(k^*) = \tau_f$. This gives,
    \begin{equation}\label{eq:kstar_st}
        k^* = k_L \St^{-1/m}.
    \end{equation}
    Thus, there is no class \rom{1} eddy for $\St \geq 1$, and no class \rom{3} eddy 
    for $\St \leq (k_L/k_\eta)^m$ (see the right panel of Figure \ref{fig:Ek_tauk}).
\item Calculate the velocity dispersion of each dust grain. The velocity
    dispersion of a dust grain induced by turbulence is given by Equation
    (6) in \citet{Markiewicz1991},
    \begin{equation}\label{eq:vp}
        \begin{split}
            v_p^2 = &\int_\rom{1} \di k E(k) (1-K^2)\\
            &+ \int_\rom{3} \di k E(k)(1-K)[g(\chi)+Kh(\chi)],
        \end{split}
    \end{equation}
    where $K=\tau_f/[\tau_f + \tau(k)]$, $g(\chi)=\chi^{-1}\arctan(\chi)$,
    $h(\chi)=1/(1+\chi^2)$ and $\chi=K\tau(k)k v_\mathrm{rel}(k)$. Here
    \rom{1} and \rom{3} denotes the integration over class \rom{1}
    ($k<k^*$) and class \rom{3} ($k \geq k^*$) eddies.
\item Calculate the cross-correlation of the velocities between grains 1
    and 2, $\langle \mathbf{v}_{p1}\cdot \mathbf{v}_{p2}\rangle$.
    From \citet{Markiewicz1991} Equation (8),
    \begin{equation}\label{eq:v1v2}
        \begin{split}
            \langle \mathbf{v}_{p1}\cdot \mathbf{v}_{p2}\rangle 
            = &\frac{1}{\tau_{f1} + \tau_{f2}}
                                  \int_{\rom{1}_{12}}\di k E(k) \\
                &\times [\tau_{f1}(1-K_{1}^2) + \tau_{f2}(1-K_{2}^2)],
        \end{split}
    \end{equation}
    where $\rom{1}_{12}$ denotes that the integration is over the eddies
    that are class \rom{1} for both grain 1 and 2, i.e., 
    $k<\mathrm{min}\{k_1^*, k_2^*\}$.
\item Obtain the collisional velocity between grains 1 and 2. Finally, the
    collisional velocity $\Delta v_{12}$ can be calculated from
    \begin{equation}\label{eq:v12}
        (\Delta v_{12})^2 = v_{p1}^2 + v_{p2}^2 
        - 2 \langle \mathbf{v}_{p1}\cdot \mathbf{v}_{p2}\rangle.
    \end{equation}
    For a given turbulence model, $(\Delta v_{12})^2$ is
    proportional to the total kinetic energy of the turbulence.
    Therefore, we usually
    present the normalized $\Delta v_{12}/v_\mathrm{tot}$ in the following
    sections.
\end{enumerate}

\section{Results}\label{section:results}
\subsection{Analytic Approximations for Grain Collisional Velocities\label{section:analytic_approximation}}
We calculate the analytic approximations of grain collisional velocities in
different regimes. The $k^*$ from Equation (\ref{eq:kstar_st}) is used to
distinguish class \rom{1} and class \rom{3} eddies.

First, we calculate $v_p^2$ in Equation (\ref{eq:vp}), which we divide into two
terms, $v_p^2 = T_\rom{1} + T_\rom{3}$,  
\begin{equation}\label{eq:T1_dvsq}
    T_\rom{1} = \int_\rom{1} \di k E(k) (1-K^2),
\end{equation}
and 
\begin{equation}\label{eq:T3_dvsq}
    T_\rom{3} = \int_\rom{3} \di k E(k)(1-K)[g(\chi)+Kh(\chi)],
\end{equation}
for the class \rom{1} and class \rom{3} eddies. 

For the $T_\rom{1}$ term, there are two possible cases: 
(1) $\St \geq 1$. There is no class \rom{1} eddy, and
$T_\rom{1}=0$. (2) $\St < 1$. In this case, for class \rom{1} eddies $\tau_f
\leq \tau(k)$, and thus we can approximate 
$1 - K^2 \approx 1 - [\tau_f/\tau(k)]^2$. This gives,
\begin{equation}\label{eq:T1}
    \begin{split}
        T_\rom{1}(\St)|_{k_L}^{k'} 
        %&=  \int_{k_L}^{k'} \di k E(k) (1 - K^2)\\
        &\approx \int_{k_L}^{k'} \di k E(k) \left[ 1 -
        \left(\frac{\tau_f}{\tau(k)}\right)^2 \right]\\
        &\approx v_\mr{tot}^2 \left[ 1 -
        \left(\frac{k_L}{k'}\right)^{p-1} \right]\\
    & - \frac{p-1}{1+2m-p} \St^2 v_\mr{tot}^2 \left[ 
        \left(\frac{k'}{k_L}\right)^{1+2m-p} - 1 \right], 
    \end{split}
\end{equation}
where 
\begin{equation}
    k' = \min\{k^*, k_\eta \}.
\end{equation}
$T_\rom{1}(\St)|_{k_L}^{k'}$ denotes the
integration of the function $E(k) (1 - K^2)$ in
the range of $k_L \leq k \leq k'$ for a grain with Stokes number $\St$.

For the $T_\rom{3}$ term, we use the approximation $g(\chi)\approx h(\chi) \approx 1$, 
following \citetalias{OC2007}. This gives
$(1-K)[g(\chi)+Kh(\chi)]\approx 1 - K^2 \approx 2\tau(k)/\tau_f$, 
with $\tau(k) < \tau_f$ for class \rom{3} eddies. There are 3 cases: (1) $\St
\geq 1$. In this case, all eddies are class \rom{3}, and 
\begin{equation}\label{eq:T3_kL_keta}
    \begin{split}
        T_\rom{3}(\St)|_{k_L}^{k_\eta} 
        %&=\int_{k_L}^{k_\eta} \di k E(k)(1-K)[g(\chi)+Kh(\chi)]\\
        &\approx \int_{k_L}^{k_\eta} \di k E(k) \frac{2\tau(k)}{\tau_f}\\
        &\approx \frac{2(p-1)}{p+m-1} \frac{v_\mr{tot}^2}{\St}  \left[ 1 -
        \left(\frac{k_L}{k_\eta}\right)^{p+m-1} \right].
    \end{split}
\end{equation}
(2) $(k_L/k_\eta)^m < \St < 1$. Class \rom{3} eddies have $k > k^*$, and
\begin{equation}
    \begin{split}
        T_\rom{3}(\St)|_{k^*}^{k_\eta} 
        %&=\int_{k^*}^{k_\eta} \di k E(k)(1-K)[g(\chi)+Kh(\chi)]\\
        &\approx \int_{k^*}^{k_\eta} \di k E(k) \frac{2\tau(k)}{\tau_f}\\
        &\approx \frac{2(p-1)}{p+m-1} \frac{v_\mr{tot}^2}{\St}  \\
        &\quad \times \left[ \left(\frac{k_L}{k^*}\right)^{p+m-1} - 
        \left(\frac{k_L}{k_\eta}\right)^{p+m-1} \right].
    \end{split}
\end{equation}
(3) $\St \leq (k_L/k_\eta)^m$. There is no class \rom{3} eddies, and $T_\rom{3}=0$.

The cross correlation term in Equation (\ref{eq:v1v2}) is only non-zero for
two grains with $\St_1 < 1$ and $\St_2 < 1$. Assuming $\St_2 < \St_1 < 1$, Equation 
(\ref{eq:v1v2}) can be written as 
\begin{equation}
    \langle \mathbf{v}_{p1}\cdot \mathbf{v}_{p2}\rangle 
    = \frac{\St_1}{\St_1 + \St_2} T_\rom{1} (\St_1)|_{k_L}^{k'_1} 
    + \frac{\St_2}{\St_1 + \St_2} T_\rom{1} (\St_2)|_{k_L}^{k'_1},
\end{equation}
where $k'_1 = \min\{k_1^*, k_\eta \}$. The analytic approximation can be
obtained from Equation (\ref{eq:T1}).

\subsection{Comparisons with Numerical Integrations\label{section:comparision_numerical}}

\begin{figure*}[htbp]
\centering
     \begin{center}
      \subfigure[Dust collisional velocities from numerical integrations.]{%
            \includegraphics[width=0.99\textwidth]{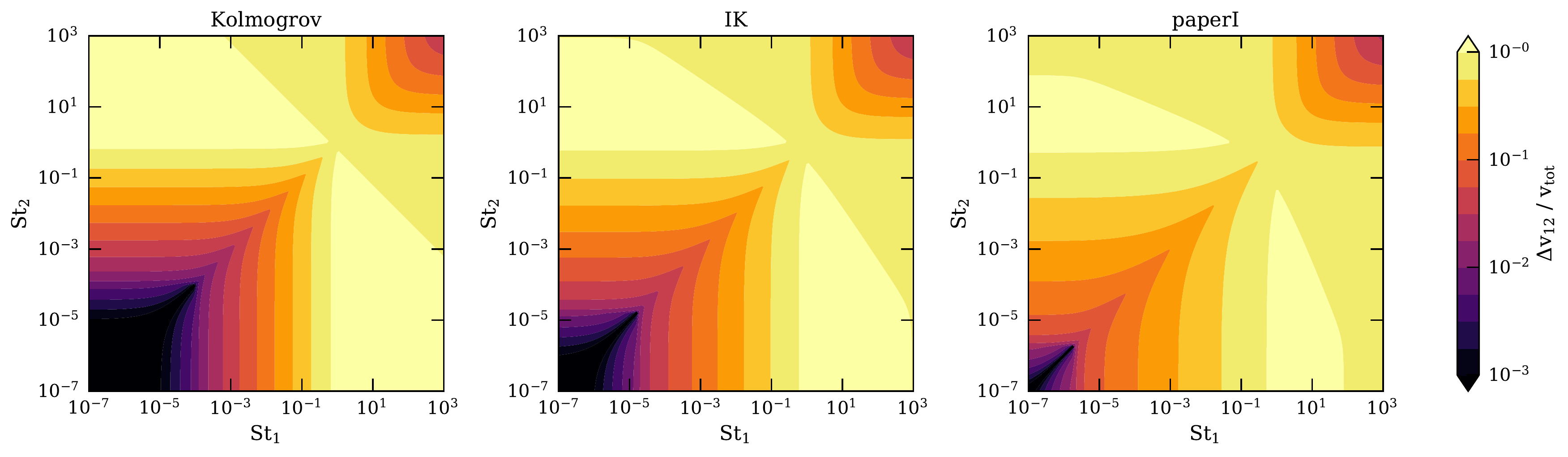}
        }
         \\
      \subfigure[Dust collisional veolcoties from analytic approximations.]{%
           \includegraphics[width=0.99\textwidth]{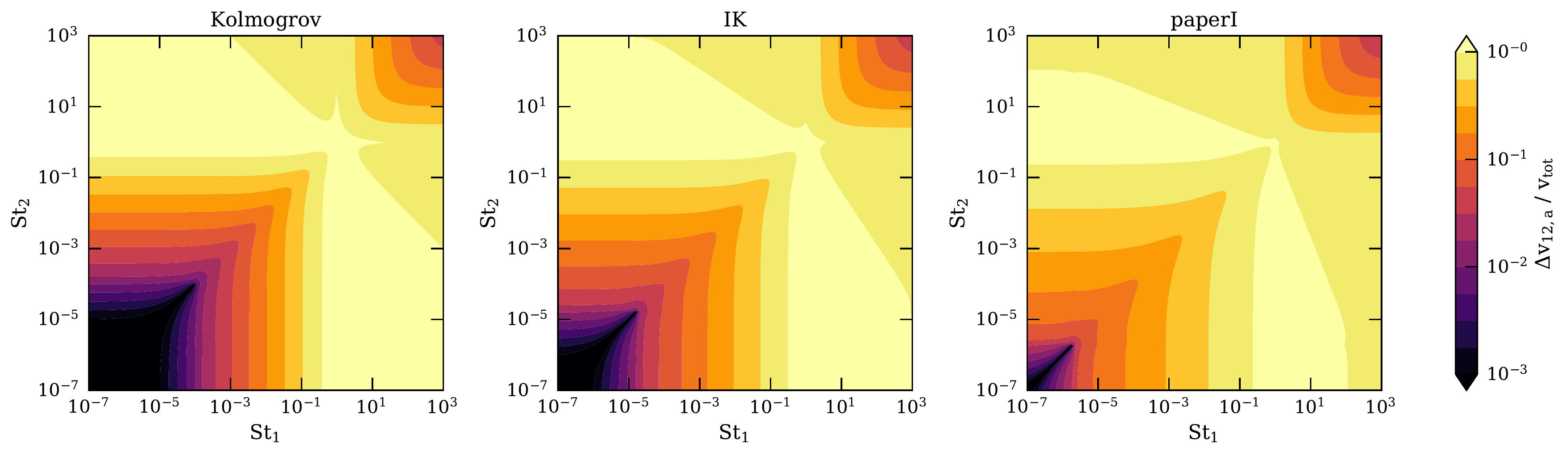}
        } 
         \subfigure[The ratio between (b) and (a).]{%
           \includegraphics[width=0.99\textwidth]{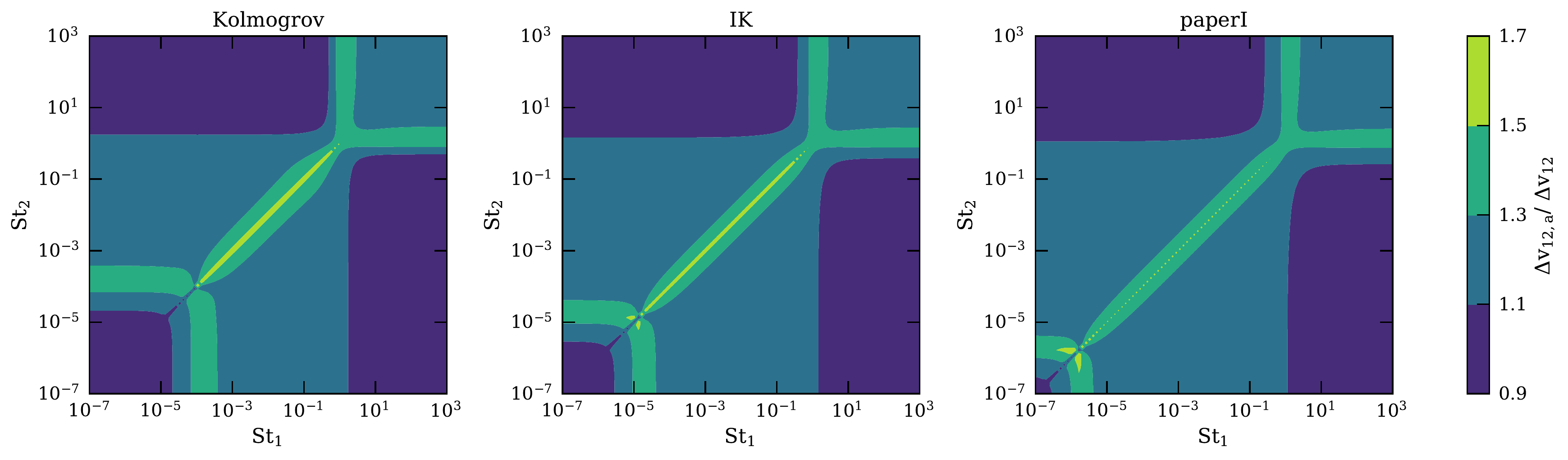}
        } 

    \end{center}
    \caption{Collisional velocities $\Delta v_{12}$ between two dust grains
    with stokes numbers $\mathrm{St_1}$ and $\mathrm{St_2}$, normalized by the
    total gas turbulent velocity $v_\mathrm{tot}$, in Kolmogorov (left), IK
    (middle) and \citetalias{dust1} (right) turbulence models. 
    (a) Collisional velocity $\Delta v_{12}$ from direct numerical integrations.  
    (b) Collisional velocity $\Delta v_{12,a}$ from analytic approximations.
    (c) The ratio $\Delta v_{12,a}/\Delta v_{12}$.
        \label{fig:vcoll}
       }
\end{figure*}

The collisional velocities $\Delta v_{12}$ between two dust grains
with Stokes numbers $\St_1$ and $\St_2$ in different turbulence
models are shown in Figure \ref{fig:vcoll}, with Reynolds number 
$\Rey=10^8$. The top panels show the collisional velocity from numerical
integrations. In the regions where $\St_1 \geq 1$ or $\St_2 \geq 1$, the
collisional velocities are very similar across different turbulence models.
However, in the regions where $\St_1, \St_2 < 1$, the collisional velocities can
differ by orders of magnitude depending on the turbulence model. This
behavior is explained in Section \ref{section:limiting_behaviors}, where we
derive the scaling relationship between $\Delta v_{12}$ and turbulence
parameters. 

The analytic approximation of $\Delta v_{12}$ is shown in the middle panels of
Figure \ref{fig:vcoll}, and the differences between the analytic approximation
and numerical integration are shown in the bottom panels. The
analytic approximation is accurate within 30\% in most regions and 
within 70\% in all regions. The largest error occurs close to $\St=1$,
$\St_1 = \St_2$, and $\St = (k_L/k_\eta)^m$, where the criteria for analytic
approximations are not satisfied (see Section
\ref{section:analytic_approximation}). 

The analytic approximation allows for fast and accurate calculation
of grain collisional velocities with arbitrary turbulence properties, without
significant sacrifice in the accuracy. The analytic formulae in Section
\ref{section:analytic_approximation} can be easily implemented in grain
growth codes, enabling the calculation of grain size evolution in
non-Kolmogorov turbulence. We provide publicly available Python scripts that
implemented our calculations 
at \url{https://github.com/munan/grain_collision}.

\subsection{Limiting Behaviors\label{section:limiting_behaviors}}
\begin{figure*}[htbp!]
\centering
\includegraphics[width=0.99\textwidth]{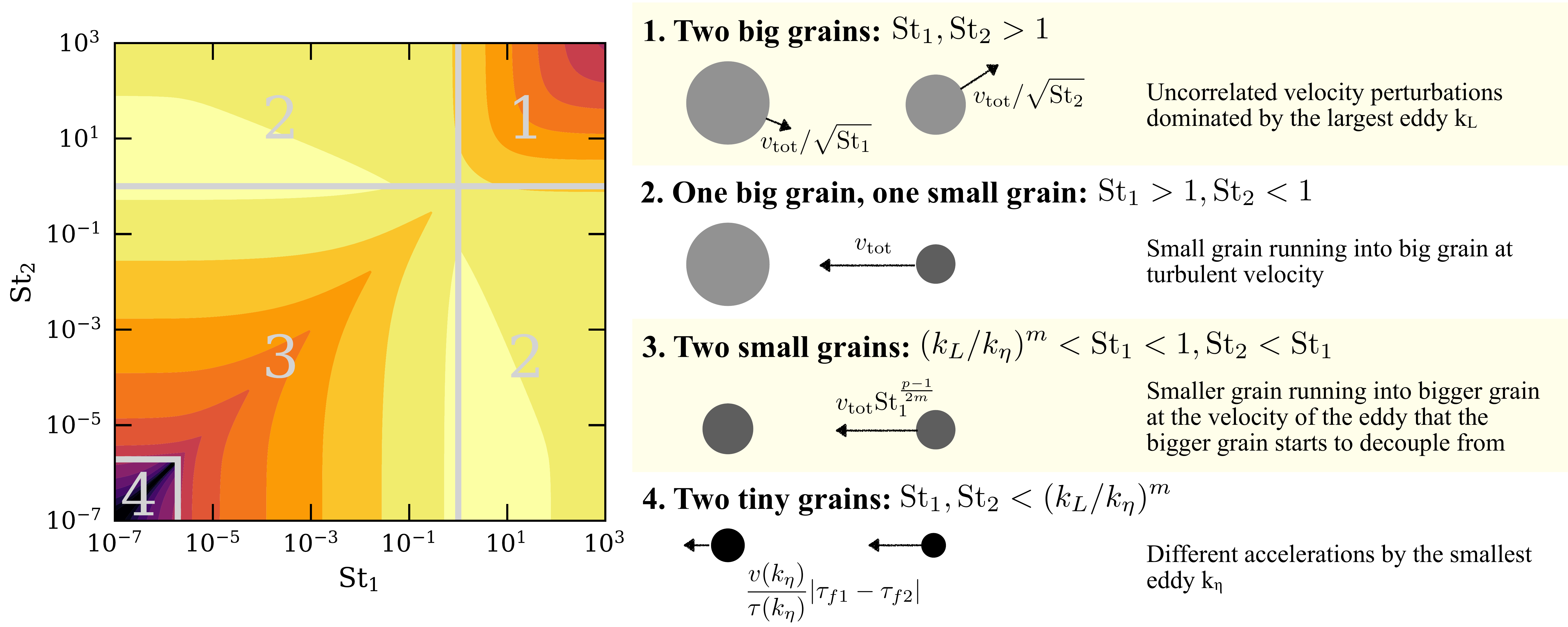}
\caption{The limiting behaviors of the grain collisional velocity $\Delta
    v_{12}$ in different regimes of the Stokes numbers.
    In regime 1 and 2, $\Delta v_{12}$ is dominated by the total turbulent
    velocity, and is insensitive to the turbulence properties. 
    In regime 3 and 4, $\Delta v_{12}$ depends
    sensitively on the turbulence properties. 
    The transition boundary between the two
    regimes and the grain collisional velocities are sensitive to the Reynolds
    number $\Rey$, the power-law slope of the turbulence energy
    spectrum $p$, 
    and power-law slope of the turbulence auto-correlation time $m$.
\label{fig:vcoll_region}}
\end{figure*}

In order to obtain a clear physical understanding of the dependence of grain
collisional velocities $\Delta v_{12}$ on the turbulence properties 
in Figure \ref{fig:vcoll}, we discuss the limiting behaviors of $\Delta v_{12}$.
We divide the stokes numbers $\St_1$ and $\St_2$ into
4 regimes and discuss the dependence of $\Delta v_{12}$ on turbulence
parameters in each regime. Figure \ref{fig:vcoll_region} summarizes the
limiting behaviors of $\Delta v_{12}$. Here we call the grains ``big'',
``small'' or ``tiny'' defined by their Stokes numbers, which determine the
scales of the turbulence eddies that they are coupled with. We always assume
that the Reynolds number is large, and therefore $k_L/k_\eta \rightarrow 0$.

\subsubsection{Two Big Grains}
Take two big grains with $\St_1 > 1$ and $\St_2 > 1$: the eddies are all class
\rom{3}, and the cross term in Equation (\ref{eq:v1v2}) vanishes.
We can use Equation (\ref{eq:T3_kL_keta})
to obtain,
\begin{equation}
    \begin{split}
        (\Delta v_{12})^2 &= v_{p1}^2 + v_{p2}^2 \\
        &= T_\rom{3}(\St_1)|_{k_L}^{k_\eta} + T_\rom{3}(\St_2)|_{k_L}^{k_\eta}\\
        &\approx \frac{2 (p-1)v_\mr{tot}^2}{p+m-1} 
        \left[1 - \left( \frac{k_L}{k_\eta} \right)^{p+m-1} \right] 
        \left( \frac{1}{\St_1} + \frac{1}{\St_2} \right)\\
        &\approx v_\mr{tot}^2 \left( \frac{1}{\St_1} + \frac{1}{\St_2} \right).
    \end{split}
\end{equation}
The last step uses the approximation that the factor $2 (p-1)/(p+m-1)$ is of order unity,
and $k_L/k_\eta \ll 1$. In this case,
each of the two grains are moving at an uncorrelated velocity of 
$v_p \approx v_\mr{tot}/\sqrt{\St}$. The
velocity perturbation is dominated by the largest eddy. 

\subsubsection{One Big Grain and One Small Grain}
Take one big grain with $\St_1 > 1$ and one small grain with $\St_2 < 1$: from
Equations (\ref{eq:T1_dvsq}) and (\ref{eq:T3_dvsq}) we have,
\begin{equation}
    \begin{split}
        (\Delta v_{12})^2 &= v_{p1}^2 + v_{p2}^2 \\
        &= T_\rom{3}(\St_1)|_{k_L}^{k_\eta} + T_\rom{1}(\St_2)|_{k_L}^{k'_2} +
    T_\rom{3}(\St_2)|_{k'_2}^{k_\eta},
    \end{split}
\end{equation}
where $k_2' = \min\{k_2^*, k_\eta \}$.
In the limit of $\St_1 \rightarrow \infty$ and $\St_2 \rightarrow 0$, the terms
$T_\rom{3}(\St_1)|_{k_L}^{k_\eta}$ and $T_\rom{3}(\St_2)|_{k'_2}^{k_\eta}$ vanish. 
Following Equation (\ref{eq:T1}),  
\begin{equation}
    (\Delta v_{12})^2 \approx T_\rom{1}(\St_2)|_{k_L}^{k'_2} \approx v_\mr{tot}^2.
\end{equation}
The bigger grain 1 barely moves due to its large mass, and the smaller grain 2 
is well-coupled with the gas and moves at the turbulent velocity $v_\mr{tot}$.

\subsubsection{Two Small Grains}
Take two small grains with $(k_L/k_\eta)^m < \St_1 < 1$ and
$\St_2 < \St_1$, we have
\begin{equation}
    \begin{split}
        (\Delta v_{12})^2 &= v_{p1}^2 + v_{p2}^2 
        - 2\langle \mathbf{v}_{p1}\cdot \mathbf{v}_{p2}\rangle\\
        &=T_\rom{1}(\St_1)|_{k_L}^{k^*_1} + T_\rom{3}(\St_1)|_{k^*_1}^{k_\eta}
        + T_\rom{1}(\St_2)|_{k_L}^{k'_2}\\
        &\quad + T_\rom{3}(\St_2)|_{k'_2}^{k_\eta}
         - 2\frac{\St_1}{\St_1 + \St_2}T_\rom{1}(\St_1)|_{k_L}^{k^*_1}\\
        &\quad -2\frac{\St_2}{\St_1 + \St_2}T_\rom{1}(\St_2)|_{k_L}^{k^*_1}.
    \end{split}
\end{equation}
We split the term $T_\rom{1}(\St_2)|_{k_L}^{k'_2}$ into two components, $T_\rom{1}(\St_2)|_{k_L}^{k^*_1} +
T_\rom{1}(\St_2)|_{k^*_1}^{k'_2}$, and neglect the second one compared to the first one, which allows us to 
approximate $T_\rom{1}(\St_2)|_{k_L}^{k'_2} \approx T_\rom{1}(\St_2)|_{k_L}^{k^*_1}$. In
addition, one can easily show that $T_\rom{3}(\St_2)|_{k'_2}^{k_\eta} <
T_\rom{3}(\St_1)|_{k^*_1}^{k_\eta}$, and therefore, the former term
can be ignored. With these approximations, we write,
\begin{equation}\label{eq:dv12_small2}
    \begin{split}
        (\Delta v_{12})^2 &\approx \frac{\St_1 - \St_2}{\St_1 + \St_2}
        [T_\rom{1}(\St_2) - T_\rom{1}(\St_1)]|_{k_L}^{k^*_1} + T_\rom{3}(\St_1)|_{k^*_1}^{k_\eta}\\
        &\approx \frac{(p-1)(\St_1 - \St_2)^2 v_\mr{tot}^2}{1+2m-p}\left[
            \left(\frac{k^*_1}{k_L}\right)^{1+2m-p} - 1 \right]\\
        &\quad + \frac{2(p-1)v_\mr{tot}^2}{(p+m-1)\St_1} 
        \left(\frac{k_L}{k^*_1}\right)^{p+m-1}. 
    \end{split}
\end{equation}
If we take the limit of $\St_2 \ll \St_1 \ll 1$, then $k^*_1/k_L = \St_1^{-1/m} \gg
1$, giving 
\begin{equation}\label{eq:v12_St}
    (\Delta v_{12})^2 \approx v_\mr{tot}^2 \St_1^{(p-1)/m}.
\end{equation}
We can understand this scaling relation by considering the coupling of the dust
grains with the gas: for eddies $k<k^*_1$, both grains are well-coupled
with the gas, and the relative velocities are small. The collisional velocity
is therefore dominated by the eddy $k^*_1$, where the larger grain 1 starts to
decouple with the gas, and the smaller grain 2 is still well-coupled with the
gas, running into the larger grain at the eddy velocity. The velocity of the
eddy $k^*_1$ is,
\begin{equation}
    v^2(k^*_1) \sim k^*_1 E(k^*_1) \sim v_\mr{tot}^2 \St_1 ^{(p-1)/m}.
\end{equation}
Here the slope of the auto-correlation time $m$ determines the scale at which the
larger grain starts to decouple, and the slope of the energy spectrum $p$
determines the eddy velocity at that scale.

\subsubsection{Two tiny Grains}
Take two tiny grains with $\St_1, \St_2<(k_L/k_\eta)^m$ and 
assuming $\St_2<\St_1$, all eddies are class
\rom{1}. Similar to Equation (\ref{eq:dv12_small2}), 
\begin{equation}
    \begin{split}
    (\Delta v_{12})^2 &\approx \frac{\St_1 - \St_2}{\St_1 + \St_2}
        [T_\rom{1}(\St_2) - T_\rom{1}(\St_1)]|_{k_L}^{k^\eta}\\
        &\approx v_\mr{tot}^2 \left( \frac{k_\eta}{k_L} \right)^{1+2m-p}
        (\St_1 - \St_2)^2\\
        &\approx \left[\frac{v(k_\eta)}{\tau(k_\eta)}(\tau_{f1} -
        \tau_{f2})\right]^2.
    \end{split}
\end{equation}
The collisional velocity is dominated by the smallest eddy $k_\eta$. The two
grains of different sizes accelerate at different rates, causing the relative
velocity. For grains with the exact same size, the collisional velocity is
zero. The collisional velocity depends on the dissipation scale $k_\eta$, as
well as the velocity and auto-correlation time of the eddy $k_\eta$. These are
in turn determined by the Reynolds number, the energy spectrum slope 
$p$ and the slope of the auto-correlation time $m$.

\section{Applications to Protoplanetary Disks}\label{section:applications}
\subsection{Grain Size Evolution\label{section:simulation}}
\begin{figure}[htbp]
\centering
\includegraphics[width=1.0\linewidth]{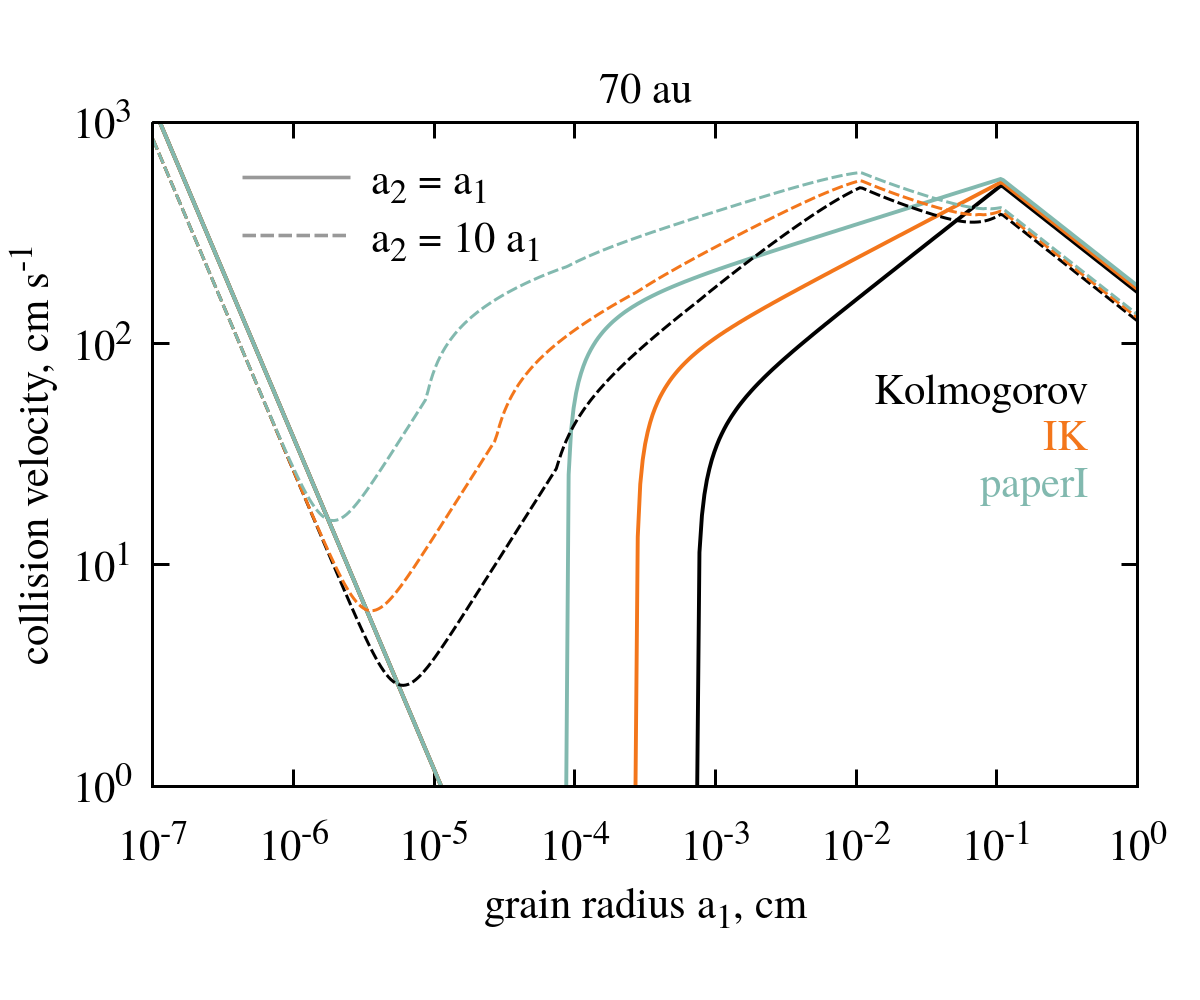}
    \caption{Illustration of collision velocities between equal-size grains (solid lines) and grains with an order of magnitude size disparity (dashed lines). The results are for our three turbulence models (Kolmogorov, IK and \citetalias{dust1}) at 70~au. For very small grains, the collisional velocity is dominated by the Brownian motion, which scales as $a^{-3/2}$. For larger grains, the collisional velocity is determined by turbulence, and can be approximately described by the 4 regimes discussed in Section \ref{section:limiting_behaviors}. Consecutive transitions between regimes 4 through 1 (see Figure \ref{fig:vcoll_region}) are evident following the dashed lines ($a_2 = 10a_1$) as $a_1$ increases. For a wide range of grain sizes from sub-micron to millimeter, the collisional velocity is very sensitive to the turbulence properties.
    \label{fig:vrel}}
\end{figure}

\begin{figure*}[htbp]
\centering
\includegraphics[width=0.45\linewidth]{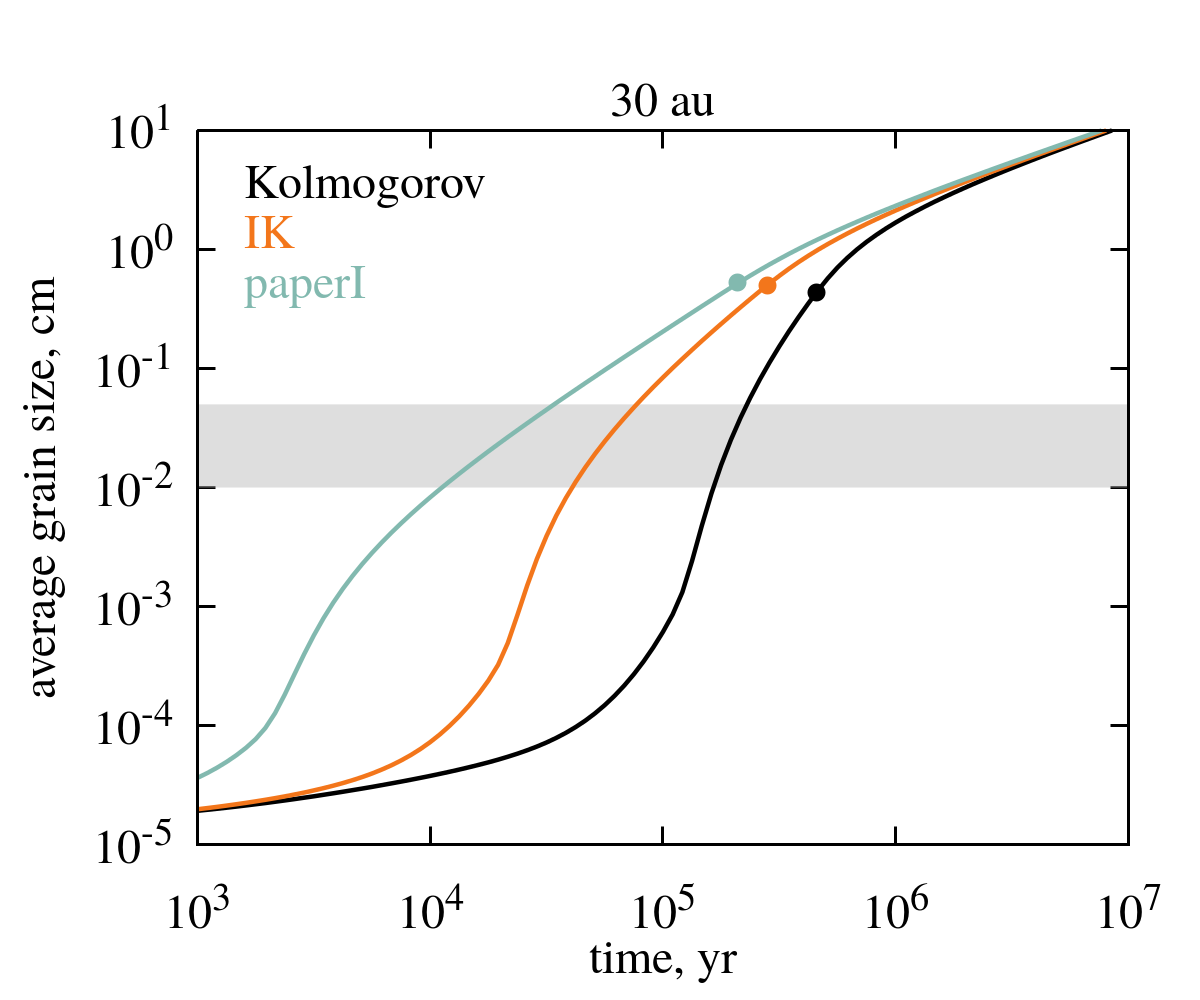}
\includegraphics[width=0.45\linewidth]{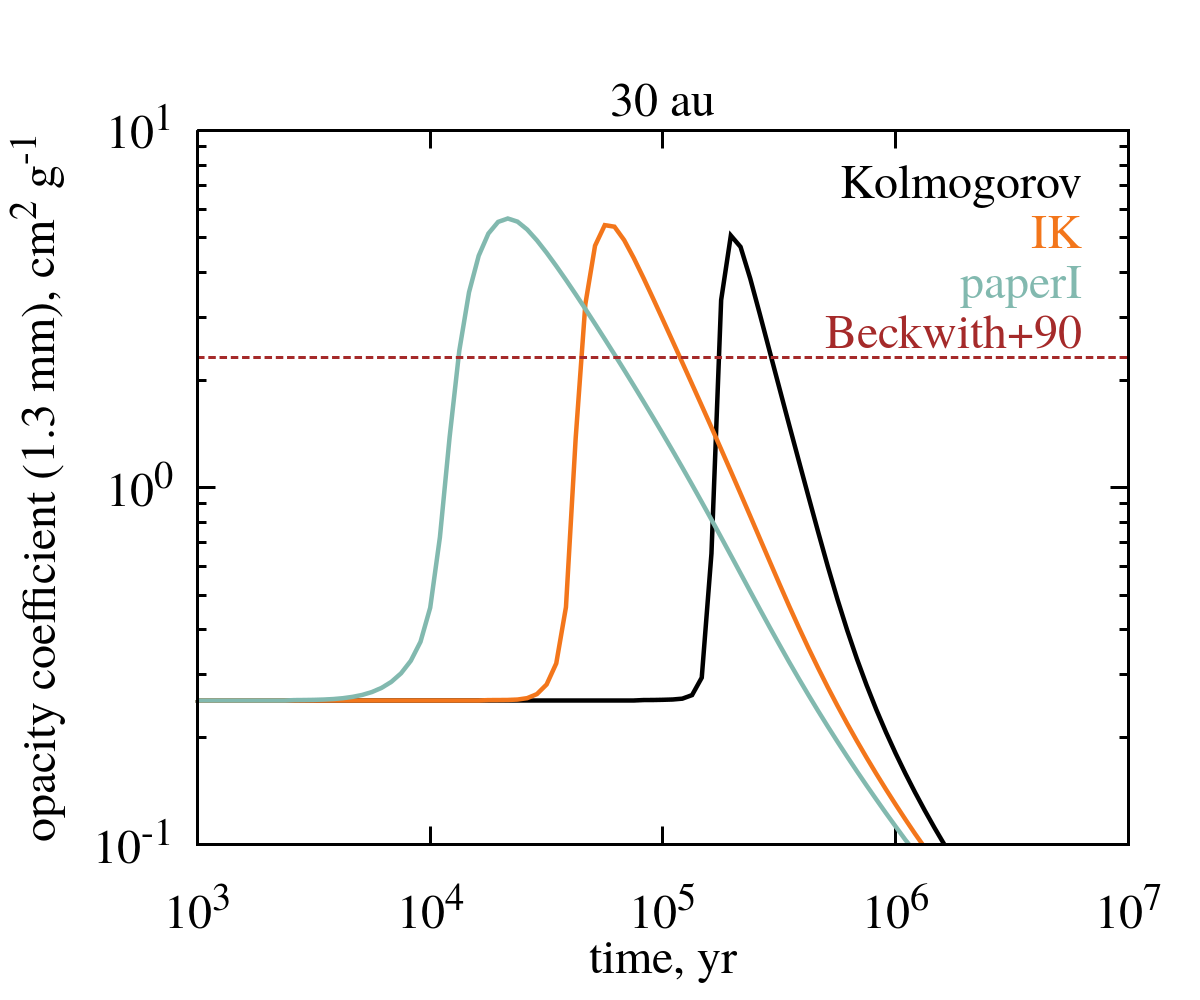}
\includegraphics[width=0.45\linewidth]{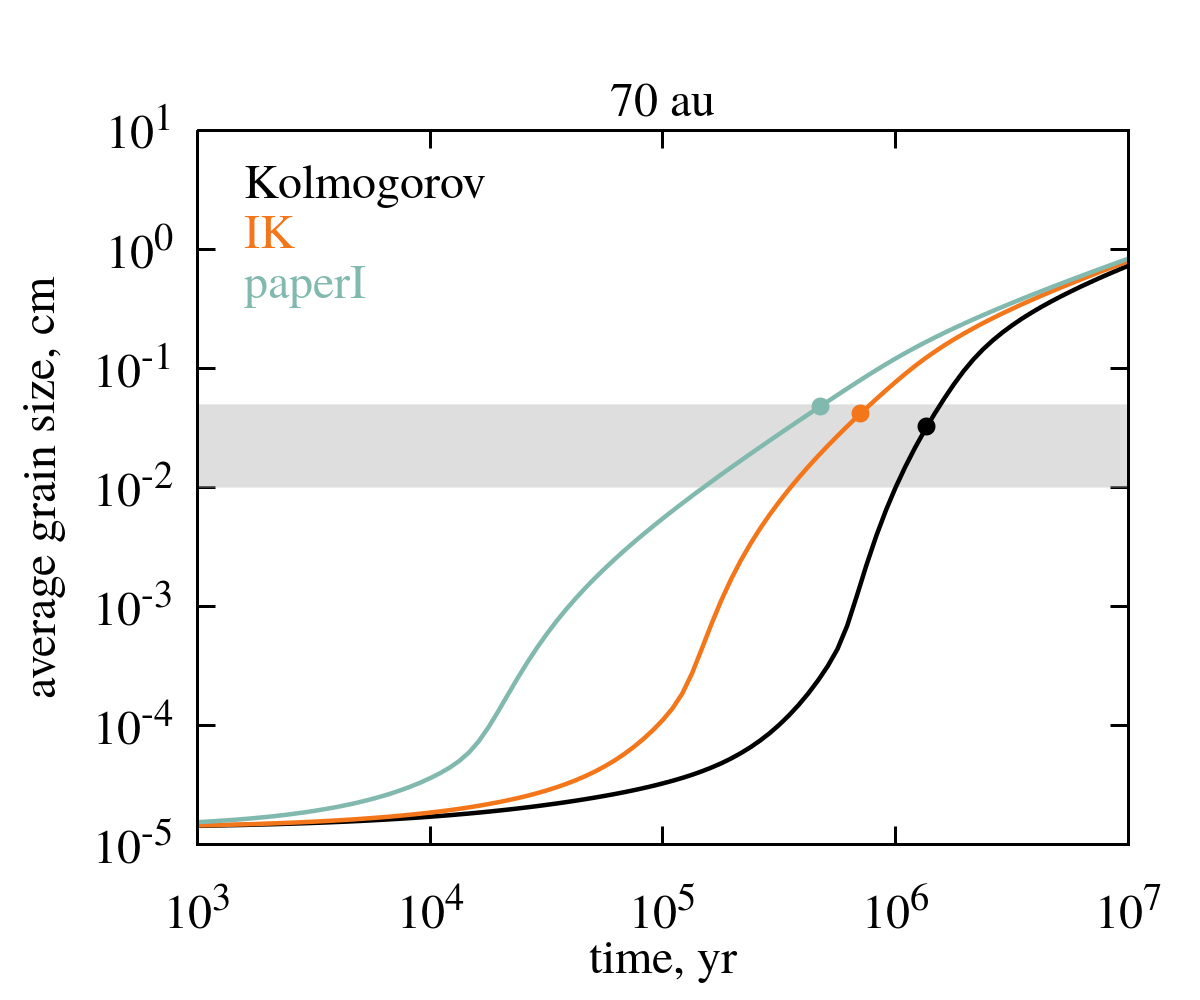}
\includegraphics[width=0.45\linewidth]{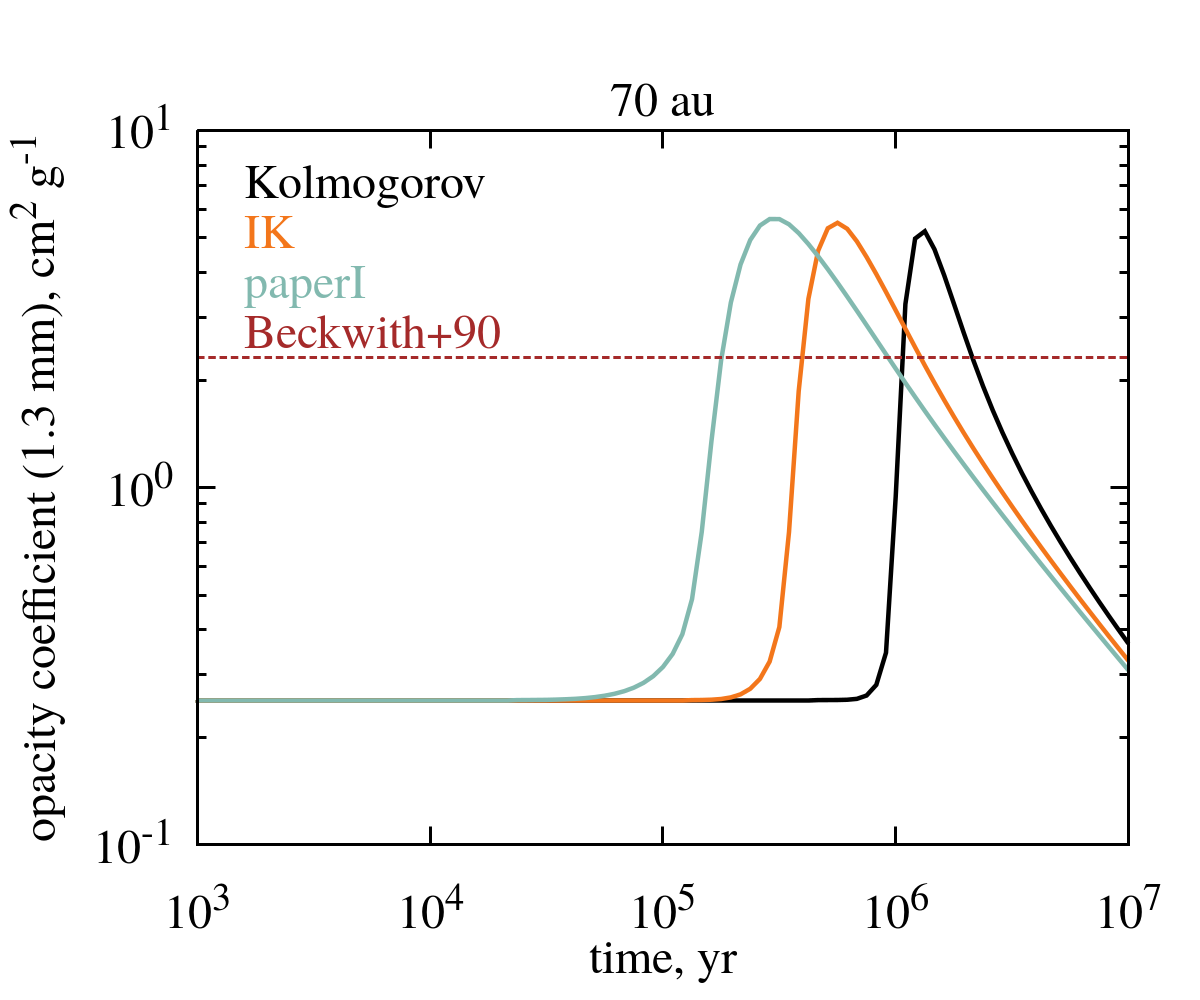}
    \caption{ Left: evolution of the average grain size (Equation \eqref{eq:adust}) in the disk mid-plane at 30 and 70~au.  
    The gray shaded region marks the size-range which contributes the most to the dust emission at mm-wavelengths. The filled circles indicate radial drift barriers for each turbulence model according to Equation \eqref{eq:a_drift}. Right: evolution of the dust opacity coefficient at the wavelength $\lambda=1.3$~mm (ALMA Band~6). The horizontal red dashed line shows the opacity coefficient of 2.3~cm$^2$~g$^{-1}$ in the observations by \citet{Beckwith1990}.
    \label{fig:amean_kappa_1e-4}}
\end{figure*}

Turbulence plays a crucial role in grain evolution in protoplanetary disks. It provides the dominant source of collisional velocities for micron- to centimeter- sized grains, which are too large to coagulate efficiently due to Brownian motion and too small to experience strong differential radial and azimuthal drift \citep{Testi2014}. 

To show the impact of turbulence properties on grain evolution, we perform numerical simulations of grain coagulation in typical disk environments, similar to \citet{Akimkin2020}. We consider the simplest case of non-charged compact spherical grains with a material density of $\rho_s=3.0$~g~cm$^{-3}$, a fixed dust-to-gas ratio of $\rho_d/\rho_g=0.01$,  and a constant turbulence 
alpha-parameter of $\alpha=10^{-4}$. This low value of $\alpha$ is motivated by recent constrains by both dust and gas observations \citep{Pinte2016, Flaherty2018}, as well as from theoretical models \citep{Simon2018}. We take two sets of physical conditions in the disk: (1) $\rho_g=5\times10^{-14}$~g~cm$^{-3}$, $T_g=38$~K; and (2) $\rho_g=3\times10^{-15}$~g~cm$^{-3}$, $T_g=25$~K, where $\rho_g$ and $T_g$ are the gas density and temperature. These correspond to the conditions in the disk mid-plane at 30 and 70~au in \citet{Akimkin2020}.  We choose to focus on the outer disk for the following reasons: (1) it is better probed by observations with its larger spacial scales and longer evolution time scales; and (2) the grain growth is less affected by fragmentation and radial drift, which we do not include in our model. In fact, for the parameter we choose, the maximum grain collisional velocity is $v_\mr{coll} \approx v_\mr{tot} \approx 4~\mr{m/s}$, smaller than the typical fragmentation velocity of $\sim 10~\mr{m/s}$ \citep{GB2015}. 
%Using Equation (33) in \citet{Birnstiel2016}, the radial drift barrier occurs for grains with sizes of $\sim 0.1$ and $\sim 0.01$ cm at the 30 and 70 au locations we considered.
Using Equation (\ref{eq:a_drift}) in Section \ref{section:a_drift}, we calculated that the radial drift barrier occurs for grains of sizes $\approx5$ mm at 30 au and $\approx0.4$ mm at 70 au (marked by the filled circles in Figure \ref{fig:amean_kappa_1e-4}).
However, we note that the radial drift is very sensitive to the disk structure commonly observed \citep{Andrews2020, Segura-Cox2020}. Non-smooth structures such as gaps and rings in the disk will significantly deter the drift. 

The initial grain size distribution is taken to be a power-law with the slope of $-3.5$ in the range of $0.005-0.25~{\rm \mu}$m \citep{MRN1977}. The coagulation equation is solved on a grid of grains masses ranging from $10^{-20}$ to $10^{7}$~g (roughly corresponding to sizes from $10^{-7}$ to $10^2$ cm) with 512 bins, providing resolution of $\sim20$ bins per grain mass decade or $\sim60$ bins per grain size decade. 

We consider two sources of grain collisional velocities: the Brownian motion $\Delta v_{\rm Br}$ and turbulence-induced velocities $\Delta v_{12}$. The total grain collisional velocity is $\Delta v_{\rm coll}=\sqrt{(\Delta v_{\rm Br})^2 +(\Delta v_{12})^2}$. In Figure~\ref{fig:vrel} we show the collisional velocities between equal-size grains and grains with an order of size disparity, calculated at 70~au. 
The Brownian motion with $\Delta v_{\rm Br}\propto a^{-3/2}$ dominates
for the smallest sizes. For $a$ near the transition to
turbulence-driven motion, the resulting collision velocity exhibits a
deep minimum, naturally leading to a slower coagulation
for these sizes. This explains the small bump seen for micron-size
grains in the size distribution (see the left panel of
Figure \ref{fig:Fig_appB} in Appendix \ref{appendix:dustcoag}).
For a large range of grain sizes from sub-micron to millimeter, the collisional velocity is very sensitive to the turbulence model, leading to dramatically different growth rates. To demonstrate the observational effect of different turbulence models on grain size evolution, we calculate the dust opacity coefficient $\kappa_{\nu}(t)=\int \pi a^2 Q_{\rm abs}(a,\nu)f(m,t)\,{\rm d}m / \rho_{\rm d}$, where $f(m,t)$ is the grain mass distribution, $\rho_{\rm d}=\int mf(m,t)\,{\rm d}m$,  and $Q_{\rm abs}(a,\nu)$ is the absorption efficiency obtained using the Mie theory for spherical silicate grains \citep{Draine1984, Akimin2020a}. 

The left panels of Figure \ref{fig:amean_kappa_1e-4} show the evolution of the average grain radius for the three cases of turbulence models (Kolmogorov, IK, and \citetalias{dust1}). The average grain radius $\bar{a}(t)$ is calculated from
\begin{equation}\label{eq:adust}
    \frac{4\pi}{3}\rho_{\rm s}\bar{a}^3(t)=\frac{\int m^2f(m,t)\,{\rm d}m}{\rho_{\rm d}}.
\end{equation}
The corresponding dust opacity at the wavelength of 1.3 mm (ALMA Band 6) is illustrated in the right panels of Figure \ref{fig:amean_kappa_1e-4}. 
The evolution of grain size distribution and dust opacity at 70 au is presented in Figure \ref{fig:Fig_appB} in Appendix \ref{appendix:dustcoag}. The evolution at 30 au is qualitatively similar to that at 70 au, but occurs faster due to the higher density and turbulent velocity.
The higher collisional velocity for smaller grains in the IK and \citetalias{dust1} turbulence makes their growth faster than in the standard Kolmogorov turbulence case.
The size range of $0.1-0.5$ mm (gray shaded region in the left panels of Figure \ref{fig:amean_kappa_1e-4}) is important, as such grains contribute the most to the disk millimeter emission \citep{Rosotti2019, Akimin2020a}. This is also shown in the right panels of \ref{fig:amean_kappa_1e-4}: as the average grain size increases, the 1.3 mm dust opacity first increases then decreases, peaking around the $0.1-0.5$ mm size-range.
At 70 au, the grains grow to sub-millimeter sizes within 0.1 Myr with the \citetalias{dust1} turbulence, while for the Kolmogorov turbulence, it takes 1 Myr to reach the same sizes. 

The faster grain growth enabled by the non-Kolmogorov turbulence has interesting implications in many observational and theoretical aspects.
This can result in rapid radial drift of dust grains in the outer disk, leading to the small disk sizes observed in class 0 and \rom{1} protostars \citep{Segura-Cox2016, Segura-Cox2018}. Furthermore, it is known that grain charging may completely stop coagulation for micron-sized grains, especially for fluffy aggregates \citep{2009ApJ...698.1122O, Akimkin2020}. The higher collisional velocities provided by the IK and \citetalias{dust1} turbulence can  help to overcome this charge barrier. Generally, the faster grain growth provides more favorable conditions for early planet formation in young disks,
by both accelerating the core formation processes and supplying solid material from the outer disk by the radial drift. 

\subsection{Fragmentation Barrier\label{section:a_frag}}
We calculate the fragmentation barrier for grain growth following \citet{Birnstiel2012}, but take into account the dependence of grain collisional velocities on turbulence properties. We estimate the collisional velocities of dust grains with $\St < 1$ from Equation (\ref{eq:v12_St}),
\begin{equation}\label{eq:vcoll_small}
    \Delta v_\mr{coll}\approx  \sqrt{\alpha} c_s \St^{\frac{p-1}{2m}},
\end{equation}
where $c_s$ is the sound speed.
By equating $\Delta v_\mr{coll}$ to the fragmentation velocity $v_\mr{frag}$, we obtain the fragmentation barrier site, 
\begin{equation}\label{eq:a_frag}
    a_\mr{frag} = \frac{\pi}{2} \frac{\Sigma_g}{\rho_s}\left[
    \frac{1}{\alpha}\left(\frac{v_\mr{frag}}{c_s}\right)^2\right]^\frac{m}{p-1}.
\end{equation}
For Kolmogorov turbulence, where $m/(p-1)=1$, we recover Equation (8) in \citet{Birnstiel2012} within a pre-factor of order unity. 
For both the IK and \citetalias{dust1} turbulence, the higher collisional velocities lead to smaller values of $a_\mr{frag}$ than that for the Kolmogorov turbulence. 
We note that the largest collisional velocity occurs for grains with 
$\St \approx 1$ (see Section \ref{section:limiting_behaviors}) at $\Delta v_\mr{coll}\approx  \sqrt{\alpha} c_s$. If $\sqrt{\alpha} c_s < v_\mr{frag}$, the fragmentation barrier is never reached (which is true for the cases considered in Section \ref{section:simulation}). 

\subsection{Radial Drift Barrier\label{section:a_drift}}
Similar to the fragmentation barrier, the radial drift barrier also depends on the turbulence properties, which influences the grain growth timescale. From \citet{Birnstiel2012}, the grain growth timescale is,
\begin{equation}
    \tau_\mr{grow} = \frac{a\rho_s}{\rho_d \Delta v_\mr{coll}},
\end{equation}
with the collisional velocity from Equation (\ref{eq:vcoll_small}). 
%We assume that the dust density $\rho_d$ is a constant ignoring vertical settling, with the dust to gas ratio $\epsilon=\rho_d/\rho_g=0.01$.
The dust density $\rho_d$ is obtained from $\rho_d = \Sigma_d/(\sqrt{2\pi}h_d)$, where the dust surface density is given by $\Sigma_d=\epsilon \Sigma_g$ with a constant dust to gas ratio $\epsilon=0.01$. The dust disk scale-height $h_d$ is calculated by \citet{YL2007},
\begin{equation}\label{eq:h_d}
    h_d = \sqrt{\frac{\alpha}{\St}} h_g,
\end{equation}
where $h_g$ is the gas disk scale-height.\footnote{The dust disk scale-height in Equation \eqref{eq:h_d} is obtained from the balance of vertical settling and turbulent diffusion of dust grains. Turbulent diffusion is dominated by the largest eddy and not sensitive to the detailed energy spectrum \citep{YL2007}. Therefore, Equation \eqref{eq:h_d} can be applied to all three turbulence models considered in this work.}
The drift timescale is,
\begin{equation}
    \tau_\mr{drift} = \frac{r v_k}{\gamma \St c_s^2},
\end{equation}
where $r$ is the disk radius and $\gamma = |\di \ln P/\di \ln r|$ is the absolute value of the power-law index of the gas pressure profile in the disk. From $\tau_\mr{grow}=\tau_\mr{drift}$, we obtain the drift barrier site,
\begin{equation}\label{eq:a_drift}
    a_\mr{drift} = \frac{2}{\pi}\frac{\Sigma_g}{\rho_s}
    \left[  \left(\frac{\pi}{8} \right)^{1/2} \frac{\epsilon}{\gamma}
    \left(\frac{h_g}{r}\right)^{-2} \right]^{\frac{2m}{3m-p+1}}.
\end{equation}
For Kolmogorov turbulence, 
\begin{equation}
    a_\mr{drift, Kol} = \frac{1}{\sqrt{2\pi}}\frac{\Sigma_d}{\rho_s \gamma}\left(\frac{h_g}{r}\right)^{-2}, 
\end{equation}
which recovers the result from Equation (33) in \citet{Birnstiel2016}. The values of $a_\mr{drift}$ are higher for the IK and \citetalias{dust1} turbulence compared to that for the Kolmogorov turbulence, due to the higher collisional velocities and faster grain growth rates (Figure \ref{fig:amean_kappa_1e-4}).
%\begin{equation}\label{eq:a_drift}
%    a_\mr{drift} = \frac{2}{\pi}\frac{\Sigma_g}{\rho_s}
%    \left[ \epsilon \left(\frac{\pi}{8} \alpha \right)^{1/2} \gamma^{-1}
%    \left(\frac{h}{r}\right)^{-2} \right]^{\frac{2m}{4m-p+1}}.
%\end{equation}

\section{Summary}\label{section:summary}
In this paper, we calculate of grain collisional velocities for an arbitrary turbulence model characterized by three dimensionless parameters: the slope of the kinetic energy spectrum $p$, the slope of the auto-correlation time $m$, and the Reynolds number $\Rey$. Our work is a significant extension of calculations by \citetalias{OC2007}, which being widely adopted in the literature, is only applicable to the Kolmogorov turbulence. As an example, we focus on three different turbulence models: the standard Kolmogorov turbulence, the IK model of MHD turbulence, and the MRI turbulence described in \citetalias{dust1}. We calculate the grain collisional velocities using numerical integration. In addition, we derive accurate analytic approximations of the collisional velocities, and give scaling relations with the Stokes numbers and turbulence properties. To demonstrate the implications, we perform numerical simulations of the grain size evolution in the outer regions of protoplanetary disks, and calculate the fragmentation and radial drift barrier for grain growth in non-Kolmogorov turbulence models. The main findings of this paper are summarized as follows:
\begin{enumerate}
    \item We calculate the grain collisional velocities between two dust grains in different turbulence models using both numerical integration and analytic approximations (Figure \ref{fig:vcoll}). The analytic approximation is simple and accurate, and can be readily implemented in complex numerical codes to model the grain size evolution in arbitrary turbulence models. We provide publicly available python scripts implementing our calculations at \url{https://github.com/munan/grain_collision}.
    \item We introduce 4 characteristic regimes for the collisional velocities, depending on the Stokes numbers of the grains.
    We perform a detailed analysis of each regime, revealing the dominant mechanism that governs the collisional velocities and presenting the corresponding 
    scaling relations (Figure \ref{fig:vcoll_region}). In particular, we show that the collisional velocities of small grains with $\St<1$ depend sensitively on the turbulence properties, with $(\Delta v_{12})^2 \sim \St^{(p-1)/m}$ (Equation \eqref{eq:v12_St}). 
    \item The collisional velocity of small grains in IK and \citetalias{dust1} turbulence are much higher (more than an order of magnitude for some grain sizes) than that in the Kolmogorov turbulence (Figure \ref{fig:vrel}). 
    \item We perform numerical simulations of grain size evolution in the outer parts of protoplanetary disks. Compared to the Kolmogorov turbulence, the higher grain collisional velocities lead to faster grain growth in the IK and \citetalias{dust1} turbulence models (Figure \ref{fig:amean_kappa_1e-4}). For the MRI turbulence in \citetalias{dust1}, grains can grow to sub-mm sizes within $\sim 0.1$ Myr even with a very low turbulence level ($\alpha=10^{-4}$) at 70 au. For Kolmogorov turbulence, growth to such sizes takes $\sim 1$ Myr. 
    \item The faster grain growth in the IK and \citetalias{dust1} turbulence may lead to rapid increase of dust opacity at mm-wavelength (Figure \ref{fig:amean_kappa_1e-4}). Increased collisional velocities can also help to overcome the charge barrier for the coagulation of micron-sized dust grains, accelerate the growth of pebbles and planetesimals, and thus promote planet formation in very young disks.
    \item We calculate the fragmentation and drift barriers for grain growth in non-Kolmogorov turbulence (Equations \eqref{eq:a_frag} and \eqref{eq:a_drift}). Compared to the Kolmogorov turbulence, the higher grain collisional velocities for the IK and \citetalias{dust1} turbulence lead to smaller values of $a_\mr{frag}$ and larger values of $a_\mr{drift}$.
\end{enumerate}
In the future, our calculations can be implemented in numerical codes to explore the effect of non-Kolmogorov turbulence on the grain size evolution in a wide range of environments.

\acknowledgments
M. Gong, A. Ivlev and P. Caselli acknowledge the support of the Max-Planck Society. M. Gong thanks Hubert Klahr for helpful discussions on this work.
V. Akimkin acknowledges the support of Ministry of Science and Higher Education of the Russian Federation under the grant 075-15-2020-780 (N13.1902.21.0039).

\appendix
\section{Eddy Class}\label{appendix:class}
The eddy classes in Equation (\ref{eq:eddy_class}) is determined both by the
auto-correlation time $\tau(k)=1/(kv(k))$ and the eddy crossing time
$\tau_\mr{cross}(k)=1/(kv_\mathrm{vel}(k))$. Below we show that
$v(k)\gtrsim v_\mathrm{rel}(k)$ for class \rom{1} eddies, and hence $\tau(k) \lesssim \tau_\mr{cross}(k)$ in this case. Therefore, the transition scale $k^*$ can be calculated from the condition $\tau(k) = \tau_f$.
There are three scenarios:
\begin{enumerate}[leftmargin=*]
    \item $\St \geq 1$. In this case $\tau_f > \tau_L$, and all eddies are
        class \rom{3}. 
    \item $(k_\eta/k_L)^{-m} < \St < 1$. We define $k_f$ to be
        the scale where $\tau(k_f)=\tau_f$, and thus $k_f/k_L = \St^{-1/m}$.
        Below we obtain that $v_\mr{rel}(k)$ is an increasing function of $k$ for $k\leq k_f$, whereas $\tau(k)$ always decreases with $k$. Therefore, for our purpose it is sufficient to show that $v(k_f) \gtrsim v_\mr{rel}(k_f)$.

        Because $\tau(k) \geq \tau_f$ at $k \leq k_f$, we can approximate
        $v_\mr{rel}(k)$ in Equation (\ref{eq:vrel}) with,
        \begin{equation}\label{eq:vrel_class}
            \begin{split}
                v_\mr{rel}^2(k_f) 
                %&= \int_{k_L}^{k_f} \di k E(k) \left(
                %\frac{\tau_f}{\tau_f + \tau(k)} \right)^2 \\
                 &\approx \int_{k_L}^{k_f} \di k E(k) \left(
                \frac{\tau_f}{\tau(k)} \right)^2 \\
                &= k_L E_L \St^2 \int_1^{k_f/k_L} x^{2m-p} \di x \\
                &\approx \frac{p-1}{2m+1-p} v_\mr{tot}^2 \St^2 \left( 
                \frac{k_f}{k_L} \right)^{2m+1-p}\\
                &= \frac{p-1}{2m+1-p}v_\mr{tot}^2 \St^{(p-1)/m}.
            \end{split}
        \end{equation}
        To approximate the integration, 
        the second last step of used the fact that $2m+1-p > 1$ for the turbulence
        models we considered (Table \ref{table:turb_model}) as well as $k_f/k_L > 1$.
        
        For $v(k)$, we have
        \begin{equation}\label{eq:vk_class}
            v^2(k_f) = 2k_f E(k_f)
             = 2k_L E_L \left( \frac{k_f}{k_L} \right)^{1-p}
             = 2(p-1)v_\mr{tot}^2 \St^{(p-1)/m}.
        \end{equation}
        
        Comparing Equations (\ref{eq:vrel_class}) and (\ref{eq:vk_class}), we
        obtain $v(k_f) \gtrsim v_\mr{rel}(k_f)$, and thus 
        $\tau(k_f) \lesssim \tau_\mr{cross}(k_f)$. This gives $k^*\approx k_f$
        and thus
        \begin{equation}\label{eq:kstar_class}
            k^* \approx k_L \St^{-1/m}.
        \end{equation}
    \item $\St \leq (k_\eta/k_L)^{-m}$. In this last case, we compare
        $v_\mr{rel}(k_\eta)$ to $v(k_\eta)$. From Equation
        (\ref{eq:vrel_class}), we have 
        \begin{equation}
            \begin{split}
                v_\mr{rel}^2(k_\eta) &\approx \frac{1}{2m+1-p} k_L E_L \St^2 \left( 
                \frac{k_\eta}{k_L} \right)^{2m+1-p}\\
                &\leq \frac{p-1}{2m+1-p} v_\mr{tot}^2 \left( 
                \frac{k_\eta}{k_L} \right)^{-(p-1)},
            \end{split}
        \end{equation}
        and similar to Equation (\ref{eq:vk_class}),
        \begin{equation}
            v^2(k_\eta) = 2(p-1)v_\mr{tot}^2 \left( \frac{k_\eta}{k_L}
            \right)^{-(p-1)}\\.
        \end{equation}
        This gives $v(k_\eta) \gtrsim v_\mr{rel}(k_\eta)$ and 
        $\tau(k_\eta) \lesssim \tau_\mr{cross}(k_\eta)$, similar to the previous
        case. In this case, all eddies are class \rom{1}.
\end{enumerate}
%The discussions above show that in all relevant cases, $\tau_\mr{cross} \lesssim
%\tau(k)$. Therefore, we can use only $\tau(k)$ to
%determine the eddy class. 
%However, this is from the assumption that
%$v_\mr{sys}$ is small. In the case that $v_\mr{sys} \ll v(k)$, the eddy class
%and $k^*$ can be modified by $v_\mr{sys}$.

\section{Grain Size Distribution and Opacity Coefficients}\label{appendix:dustcoag}
\begin{figure}[htbp]
\centering
\includegraphics[width=0.49\linewidth]{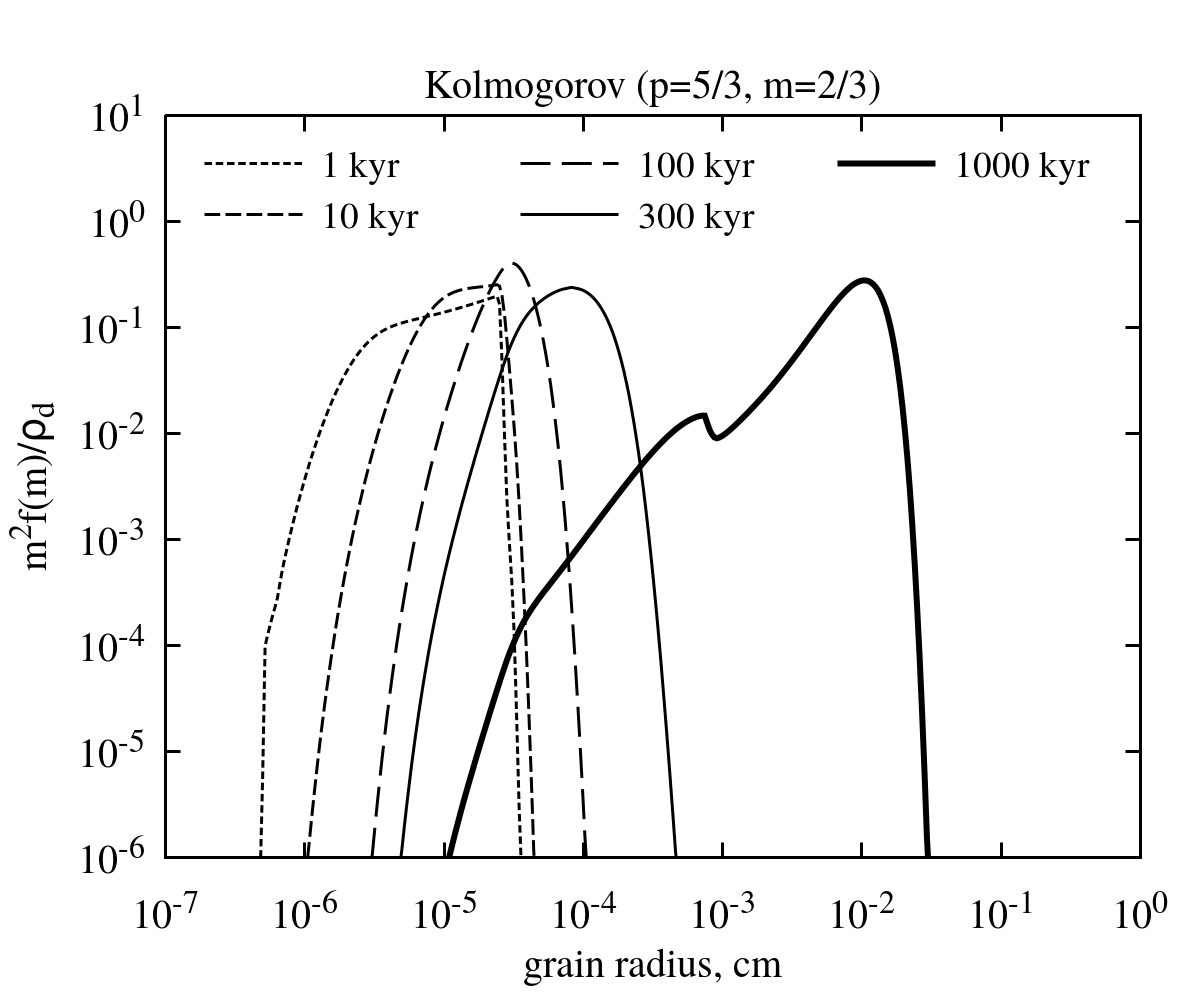}        
\includegraphics[width=0.49\linewidth]{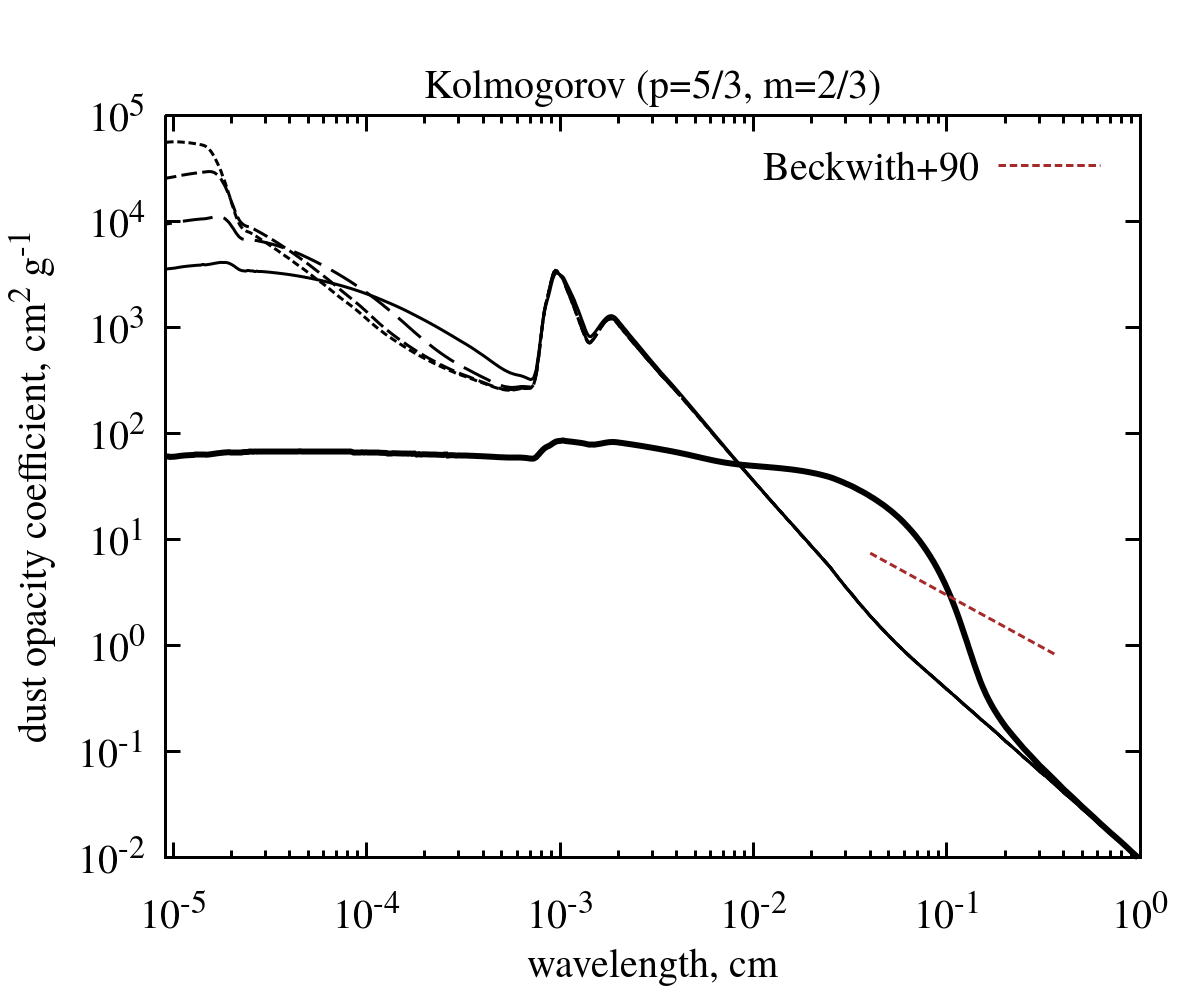}
\includegraphics[width=0.49\linewidth]{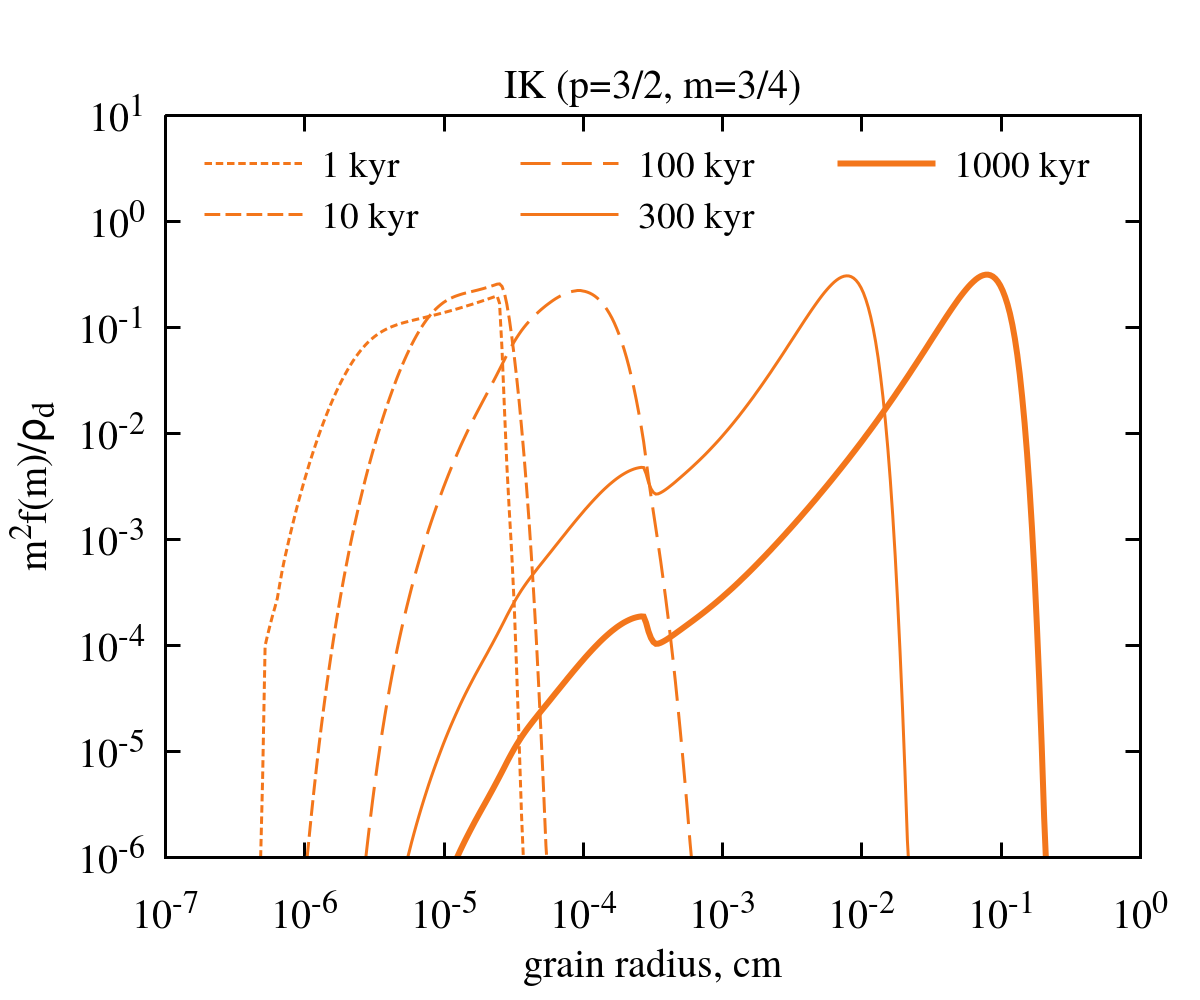}        
\includegraphics[width=0.49\linewidth]{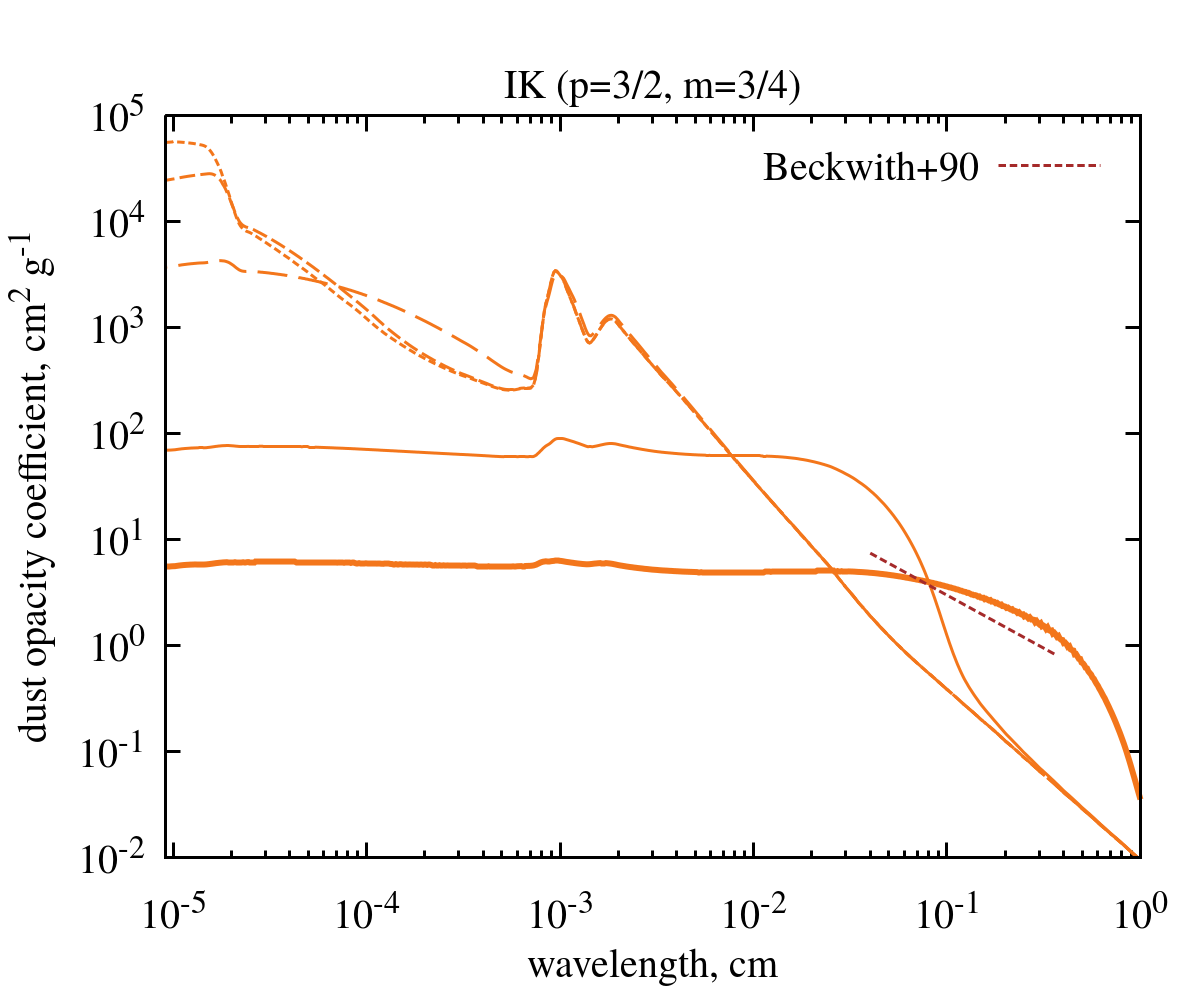}
\includegraphics[width=0.49\linewidth]{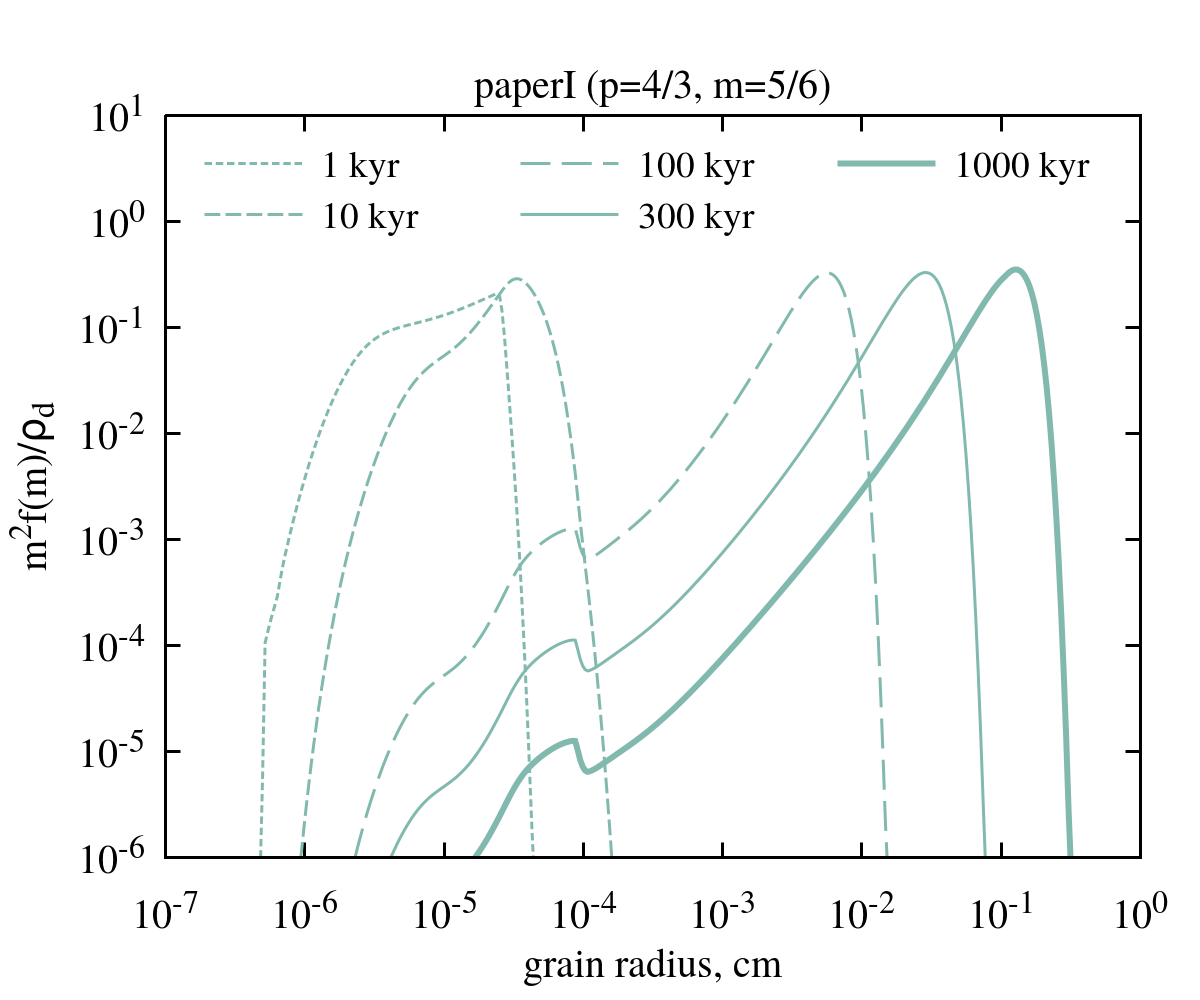}        
\includegraphics[width=0.49\linewidth]{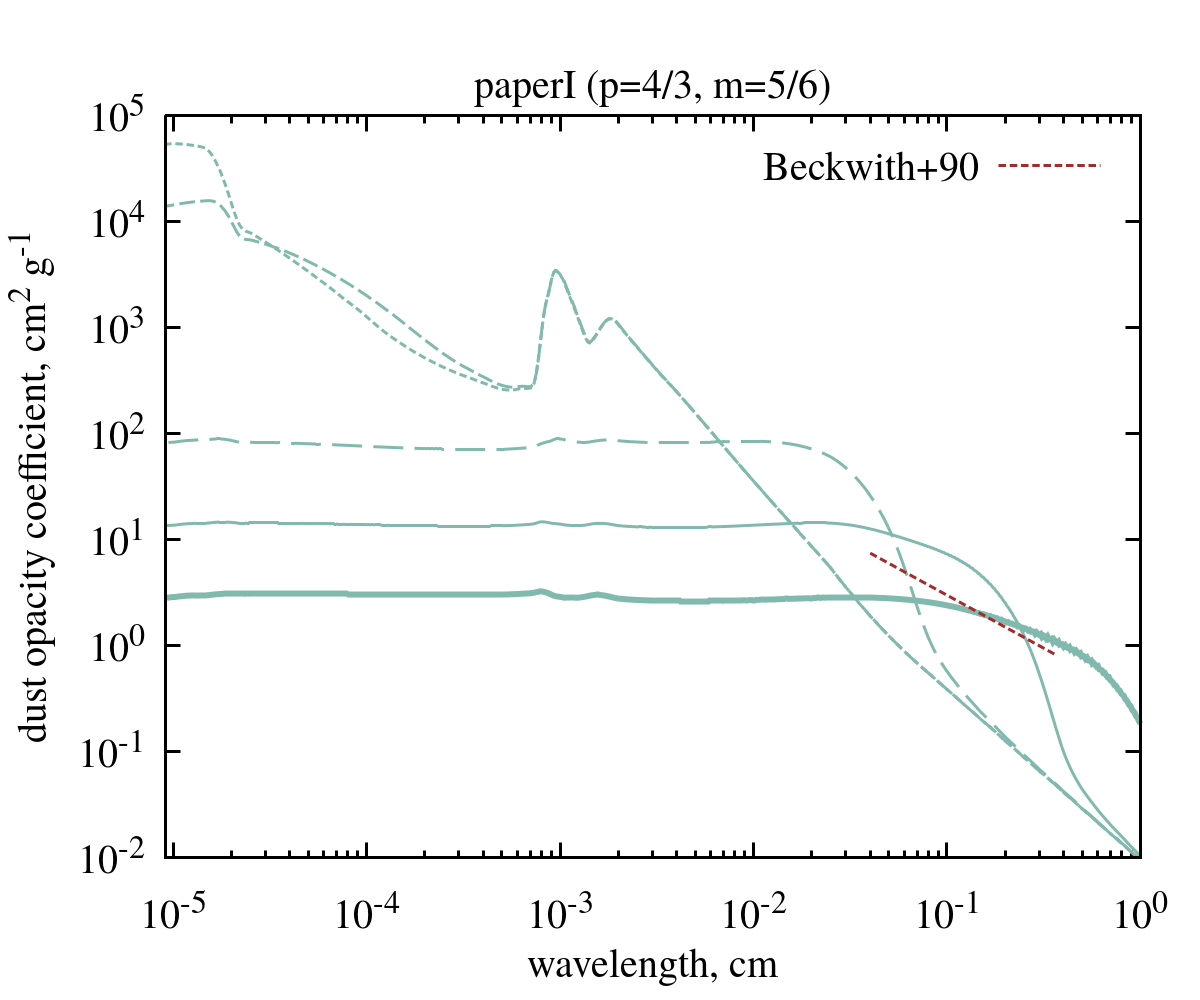}
    \caption{Evolution of the grain size distribution (left) and the corresponding opacity coefficients (right) in the disk mid-plane at 70~au. The Kolmogorov, IK and \citetalias{dust1} turbulence cases are shown in the top, middle and bottom panels. The red dashed lines in the right panels indicate the the opacity coefficients in the mm-wavelength range by \citet{Beckwith1990}.
    \label{fig:Fig_appB}}
\end{figure}
Figure \ref{fig:Fig_appB} shows the evolution of the grain size distribution and wavelength-dependent dust opacity coefficients at 70 au for the Kolmogorov, IK and \citetalias{dust1} turbulence models. The early evolution at $t\lesssim 10^4~\mr{yr}$ is governed by the Brownian motion, and therefore insensitive to the turbulence properties. Higher turbulence-induced collisional velocities in the \citetalias{dust1} case (see Figure \ref{fig:vrel}) push the peak of the grain size distribution to $\sim 100~{\rm \mu}$m already at $10^5$~yr, giving rise to an earlier increase of dust opacities in the millimeter wavelength range.

\bibliographystyle{apj}
\bibliography{apj-jour,all}
\end{document}